\documentclass[conference]{IEEEtran}

\usepackage{tikz}
\usepackage{amsmath}
\usepackage{multicol}
\usepackage{multirow}
\usepackage{graphicx}
\usepackage{subcaption}
\usepackage{booktabs}
\usepackage{caption}
\usepackage{subcaption}
\usepackage{amsfonts}
\usepackage{makecell}
\usepackage{xcolor}

\usepackage{adjustbox,mdframed}
\usepackage{xspace}
\usepackage{marginnote}

\PassOptionsToPackage{hyphens}{url}\usepackage{hyperref}
\usepackage{bookmark}
\usepackage{colortbl}
\usepackage[labelformat=simple]{subcaption}

\usepackage{amsthm}
\newcommand{\ra}[1]{\renewcommand{\arraystretch}{#1}}
\newcommand{\model}[0] {\textit{DRAINCLoG}}
\newcommand{\firstmodule}[0] {\textit{NFT Transaction Context Extractor}}
\newcommand{\secondmodule}[0] {\textit{Social Context Extractor}}

\usepackage{soul}
\def\BibTeX{{\rm B\kern-.05em{\sc i\kern-.025em b}\kern-.08em
    T\kern-.1667em\lower.7ex\hbox{E}\kern-.125emX}}

\newif\ifdiff

\newcommand{\diff}[1]{\ifdiff{\color{blue}\fi#1\ifdiff}\fi}
\newcommand{\diffnote}[1]{\ifdiff\marginnote{{\small\color{olive}\textmd{{[}#1{]}}}}[-0.25cm]\fi}

\newcommand{\diffnotenext}[1]{\ifdiff\marginnote{{\small\color{olive}\textmd{{[}#1{]}}}}[+0cm]\fi}
\difffalse
\begin{document}

\title{DRAINCLoG: Detecting Rogue Accounts with Illegally-obtained NFTs using Classifiers Learned on Graphs}

\author{
    {\rm Hanna Kim\textsuperscript{1}} \hspace{1em} 
    {\rm Jian Cui\textsuperscript{2}} \hspace{1em} 
    {\rm Eugene Jang\textsuperscript{3}} \hspace{1em} 
    {\rm Chanhee Lee\textsuperscript{3}} \hspace{1em} 
    {\rm Yongjae Lee\textsuperscript{3}} \hspace{1em} 
    {\rm Jin-Woo Chung\textsuperscript{3}} \hspace{1em} 
    {\rm Seungwon Shin\textsuperscript{1}} \hspace{1em} \texorpdfstring{\\}
    \textsuperscript{1} KAIST, Daejeon, South Korea \texorpdfstring{\\}
    \textsuperscript{2} Indiana University Bloomington, Indiana, USA \texorpdfstring{\\}
    \textsuperscript{3} S2W Inc., Seongnam, South Korea\texorpdfstring{\\}
    \textsuperscript{1}\texorpdfstring{\texttt{\{gkssk3654, claude\}@kaist.ac.kr}\\}
    \textsuperscript{2}\texorpdfstring{\texttt{\{cuijian\}@iu.edu}\\}
    \textsuperscript{3}\texttt{\{genesith, leemember, lee, jwchung\}@s2w.inc}
}

%don't want date printed
\date{}

\IEEEoverridecommandlockouts
\makeatletter\def\@IEEEpubidpullup{6.5\baselineskip}\makeatother
\IEEEpubid{\parbox{\columnwidth}{
    Network and Distributed System Security (NDSS) Symposium 2024\\
    26 February - 1 March 2024, San Diego, CA, USA\\
    ISBN 1-891562-93-2\\
    https://dx.doi.org/10.14722/ndss.2024.24888\\
    www.ndss-symposium.org
}
\hspace{\columnsep}\makebox[\columnwidth]{}}


\maketitle
\pagestyle{plain}
\begin{abstract}
% A Non-Fungible Token (NFT) is a unique and non-interchangeable cryptographic asset on a blockchain.
% As Non-Fungible Tokens (NFTs) continue to grow in popularity, NFT users have become targets of phishing attacks by cybercriminals, called \textit{NFT drainers}.
As Non-Fungible Tokens (NFTs) continue to grow in popularity, NFT users have become targets of phishing attacks by cybercriminals, called \textit{NFT drainers}.
Over the last year, \$100 million worth of NFTs were stolen by drainers, and their presence remains a serious threat to the NFT trading space.
However, no work has yet comprehensively investigated the behaviors of drainers in the NFT ecosystem.

In this paper, we present the first study on the trading behavior of NFT drainers and introduce the first dedicated NFT drainer detection system.
\diff{We collect 127M NFT transaction data from the Ethereum blockchain and 1,135 drainer accounts from five sources for the year 2022.
We find that drainers exhibit significantly different transactional and social contexts from those of regular users. 
\diffnote{R3C1}With these insights, we design \model{}, an automatic drainer detection system utilizing Graph Neural Networks. 
This system effectively captures the multifaceted web of interactions within the NFT space through two distinct graphs: the NFT-User graph for transaction contexts and the User graph for social contexts. 
Evaluations using real-world NFT transaction data underscore the robustness and precision of our model.
\diffnote{MR1} Additionally, we analyze the security of \model{} under a wide variety of evasion attacks. }

\end{abstract}
\section{Introduction}
\label{sec:intro}
With the emergence of blockchain technology, \textit{Non-Fungible Tokens} (NFTs) have revolutionized the digital creator economy. 
NFTs are digital assets, such as art or collectibles, with unique identification codes and metadata~\cite{ethereum_nft}.
NFTs have attracted numerous content creators and investors, and by 2021 the NFT market exploded, growing to around \$22 billion~\cite{nftgo}. 

% As a result, phishing scams have also emerged in the NFT ecosystem~\cite{elliptic_ana, nft_theft}. 
As a result, scammers targeting NFTs have also emerged in the NFT ecosystem~\cite{elliptic_ana, nft_theft}. 
According to blockchain analysis provider Elliptic, over \$100 million worth of NFTs were stolen by NFT phishing scammers, called \textit{drainers}, in a one-year period from July 2021 to July 2022~\cite{elliptic_ana}.
NFT drainers commonly use NFT phishing scams to obtain NFTs illegally which continue to make headlines, causing significant damage to users~\cite{scam1, scam2, scam3}.
% NFT phishing scams continue to make headlines, causing significant damage to users~\cite{scam1, scam2, scam3}. 
For instance, a sophisticated phishing attack on Uniswap\footnote{uniswap.com} (the largest Ethereum-based decentralized exchange) caused \$8 million worth of damages to NFT users~\cite{uniswap1, uniswap2}. 

Efforts to combat NFT drainers have had limited effectiveness. 
The Opensea NFT marketplace implemented the policy of marking stolen NFTs as untradeable~\cite{opensea_policy}. 
However, this is only effective when victims are able to notice and report the attacker.
The cryptowallet Metamask implemented phishing warnings~\cite{metamask1, metamask2}, but were able to be bypassed by certain drainers~\cite{drainer_update}. 

Even before NFT drainers, phishing attacks targeting cryptocurrency were already considered a significant threat to the trading security of Ethereum~\cite{chen2020survey}.
As a response to such attacks, researchers have proposed several methods to detect phishers. 
One line of research relies on using hand-crafted user features to detect scammers~\cite{chen2020phishing}, while other approaches employ network representation learning by utilizing Node2Vec~\cite{wu2020phishers} or Graph Neural Networks (GNN)~\cite{EGCN, li2022ttagn, zhang2021blockchain} to capture scammers.

However, these methods of detecting cryptocurrency phishers are unsuitable for detecting NFT drainers for the following reasons. First, features that are essential for cryptocurrency phisher detection, such as those that describe the liquidation process, are not shown in NFT drainers. In addition, cryptocurrency-focused detection systems cannot leverage individual NFTs and fail to account for the multiple transaction types of NFTs. Thus, they cannot fully capture the differences between NFT drainers from regular users. 
% (See Section \ref{sec:pre} for a detailed explanation.)
This raises the need for an automatic detection system that captures various factors of the NFT ecosystem.
This also motivates a comprehensive investigation on the transaction patterns of NFT drainers.

% To the best of our knowledge, the existing literature has not explored phishing scams in the NFT ecosystem.
To the best of our knowledge, the existing literature has not explored detecting \textit{NFT drainers} (NFT stealers).
To fill this gap, we aim to investigate the trading traits of \diff{these} drainers and propose an automatic detector that identifies suspicious trading behaviors. To this end, we collect \diff{an extensive dataset of} over 127 million NFT transactions \diff{from the Ethereum blockchain, spanning from January to December 2022}. We also collect information \diff{on} 1,135 reported drainers from \diff{various sources}, including Twitter~\cite{twitter} and Etherscan~\cite{etherscan}. 
\diff{Our analysis of these drainer accounts in the NFT ecosystem reveals their trading patterns to be distinctly different from regular users.
Specifically, we pinpoint two crucial factors in identifying drainers: their unique \textbf{NFT transaction context}, characterized by quickly selling NFTs at much lower prices, and \textbf{social context}, often linking with other users displaying similar trading anomalies.}
% (See Section \ref{sec:measurement} for more details.)

However, capturing the intricate relationships between users and NFTs for drainer detection remains a challenging task.
This is because the transaction history of millions of NFTs and the various types of interactions between millions of users must be considered.
To address this challenge, we propose a novel GNN-based drainer detection system, \model{}. We use GNNs as they are well-suited for modeling relationships between users and NFTs in a graph structure. GNNs can incorporate both the features of nodes and edges, making them effective at identifying patterns of anomalous behavior among interconnected entities~\cite{zhang2022efraudcom, akoglu2013anomaly, tan2021graph, cui2022meta}. The GNN-based structure allows \model{} to capture the user-to-user and NFT-to-user relationships effectively.

\diff{\model{} utilizes two types of graphs that uniquely model relationships to identify drainers}: the \textit{NFT-User graph} and the \textit{User graph}. 
The NFT-User graph models interactions between users and NFTs, with two types of nodes and attributed edges, allowing us to obtain a representation of users' transaction context by considering each NFT's transaction history.
The User graph models interactions between users, with attributed nodes and two types of edges, enabling us to capture the social context by integrating information on the users and their relationships. 
\diffnote{R3C1}\diff{To obtain representations, we use customized GNNs for each NFT graph type. These representations are then fused to leverage information from both graph types. Our model significantly outperforms existing baselines in drainer detection. 
}
% We also demonstrate the robustness of our model by simulating evasion attacks. 
Overall, we believe that our study will inspire further research and practical efforts to improve NFT trading security.
In summary, the contributions of the work are listed below:
\vspace{-0.2cm}
\begin{itemize}

    \item We collect 127M NFT transaction data from the Ethereum blockchain and 1,135 drainer accounts from five different channels. Drainer accounts are publicly available for future research~\footnote{will be made available on acceptance}.
    \item We present the first empirical study on NFT drainers and find that drainers have distinct characteristics and transaction patterns from regular users.
    \item \diffnote{R3C1}\diff{We comprehensively capture the relationships between users and NFTs into novel graph structures. Our GNN-based model, \model{}, automatically detects drainers from these graphs.} 
    % \item We verify the robustness of \model{} by simulating evasion attacks.
\end{itemize}

%NFT crypto 차이점은 background 보시고 아직 그 이해가 부족하다. 
%NFT 탐지 그전에 나온거는 ~~해서 한계가 있음 
%그래서 이런 새로운 시스템이 필요함 

%그래프 관련된걸 특징 분석에 대한 결과를 추가적으로 써라 -시스템이랑 이어지도록 , attribute을 쓴 이유에 대해 잘 설명되도록 
\section{Background and Motivation}
\label{sec:pre}
\subsection{Background}
\label{s:background}
% In this section, we outline the required background related to NFTs and phishing attacks in the NFT ecosystem. Then, we detail the motivation of our study.

\noindent\textbf{Non-Fungible Token}
% Ethereum is a distributed computing platform powered by blockchain technology for building distributed applications (dApps) and decentralized organizations~\cite{ethereum}. 
% There are two types of accounts in Ethereum: Externally Owned Accounts (EOA) and Contract Accounts.
% An EOA is a wallet account controlled by private keys and can initiate transactions.
% A contract account has an associated code deployed to the blockchain, called a \textit{smart contract}.
% Transactions from an EOA to a contract account can execute the code, leading to many different actions (running functions), such as transferring tokens~\cite{ether_account}.
% Ether (ETH) is the native cryptocurrency of Ethereum created by the underlying Ethereum protocol, not by users.
% Unlike Ether, tokens can be for various transactions but are usually used to represent an asset.
% tokens can be created by developers by deploying a type of smart contract called \textit{token contracts}. Tokens can be used for various transactions, but are usually used to represent an asset.
% There are two main executed standards for token contracts in Ethereum: \textit{ERC-20} and \textit{ERC-721}. ERC-20~\cite{erc20} introduces a standard for Fungible Tokens (FTs) that are interchangeable with other tokens of the same type.
% On the other hand, ERC-721~\cite{erc721} is a standard for the issuance and trading of non-fungible assets, called \textit{Non-Fungible Tokens}.
(NFT) is a cryptographic asset on the blockchain with unique identification codes and metadata that distinguish them from each other. 
NFTs guarantee ownership of unique digital assets, such as images, video files, and even physical assets.
\diff{By 2017, the Ethereum blockchain, which currently accounts for approximately 80\% of the global NFT trading volume, became a popular hub for NFTs, with collectibles like CryptoPunks~\cite{nft_stat} gaining prominence.}
An NFT \textit{collection} \diff{refers to a group} of NFTs \diff{sharing similar features, but each NFT has unique variations that sets it apart.
This uniqueness can lead to significant value differences among NFTs, even if they belong to the same collection.}
% As an example, Cryptopunks, one of the earliest and most popular collections, is a collection of 10,000 uniquely generated characters. Most characters are of either the Male or Female type, but only a few characters are of rarer types (i.e., Alien, Ape, or Zombie)~\cite{cryptopunks}.
% Most popular NFT collections have their own platform for trading. 

\begin{figure}
    \centering
    \includegraphics[width=0.95\linewidth]{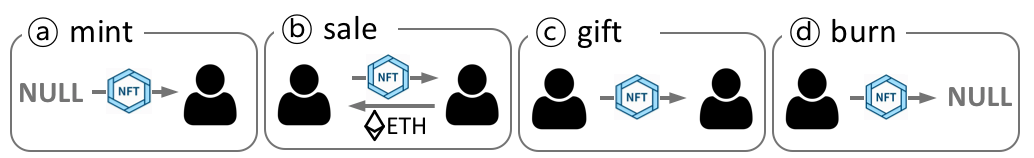}
    \caption{Summary of NFT transaction types: \textcircled{\small a} Mint, \textcircled{\small b} Sale, \textcircled{\small c} Gift, and \textcircled{\small d} Burn}
    \label{f:nft_txtype}
    \vspace{-0.4cm}
\end{figure}

\noindent\textbf{NFT Transaction Types} 
We introduce four types of NFT transactions: \textit{mint, sale, gift,} and \textit{burn}, as depicted in Figure~\ref{f:nft_txtype}.

\textcircled{\small a} \textit{Mint}. An NFT is created by minting, the process of inscribing a digital asset to the blockchain.
Minted NFTs can be listed and traded on NFT marketplaces, such as OpenSea~\cite{opensea} and Rarible~\cite{rarible}.

\textcircled{\small b} \textit{Sale}.
A sale is a process of transferring an NFT ownership to another account for payment.
NFTs are typically traded with Ether, the native cryptocurrency of Ethereum, and sometimes fungible tokens. 
Users can partake in sales in two ways: buying and selling.

\textcircled{\small c} \textit{Gift}.
A gift is a process of handing over ownership without any \diff{monetary exchange}. 
\diff{Typically, gifting occurs between addresses that are related.}
\diff{Within the NFT ecosystem, there are various scenarios where NFTs are gifted. For example, gifts can be used between users to avoid monitoring when manipulating markets by wash trading}
Users can partake in gift transfers in two ways: gifting-in and gifting-out.

\textcircled{\small d} \textit{Burn}. An NFT cannot be deleted from the blockchain, but it can be ‘burned’. Burning is the process of sending NFTs to an inaccessible address (NULL), which will remove them from circulation. 
Burning is used for various purposes, such as adjusting the supply of NFTs, operating a collection's community, etc.

% We also divide transactions into \textit{in-transaction} and \textit{out-transaction}, depending on the direction. In-transactions include buying, gifting-in, and minting. Out-transactions include selling, gifting-out, and burning.

\begin{figure}[t]
    \centering
    \includegraphics[width=0.37\textwidth]{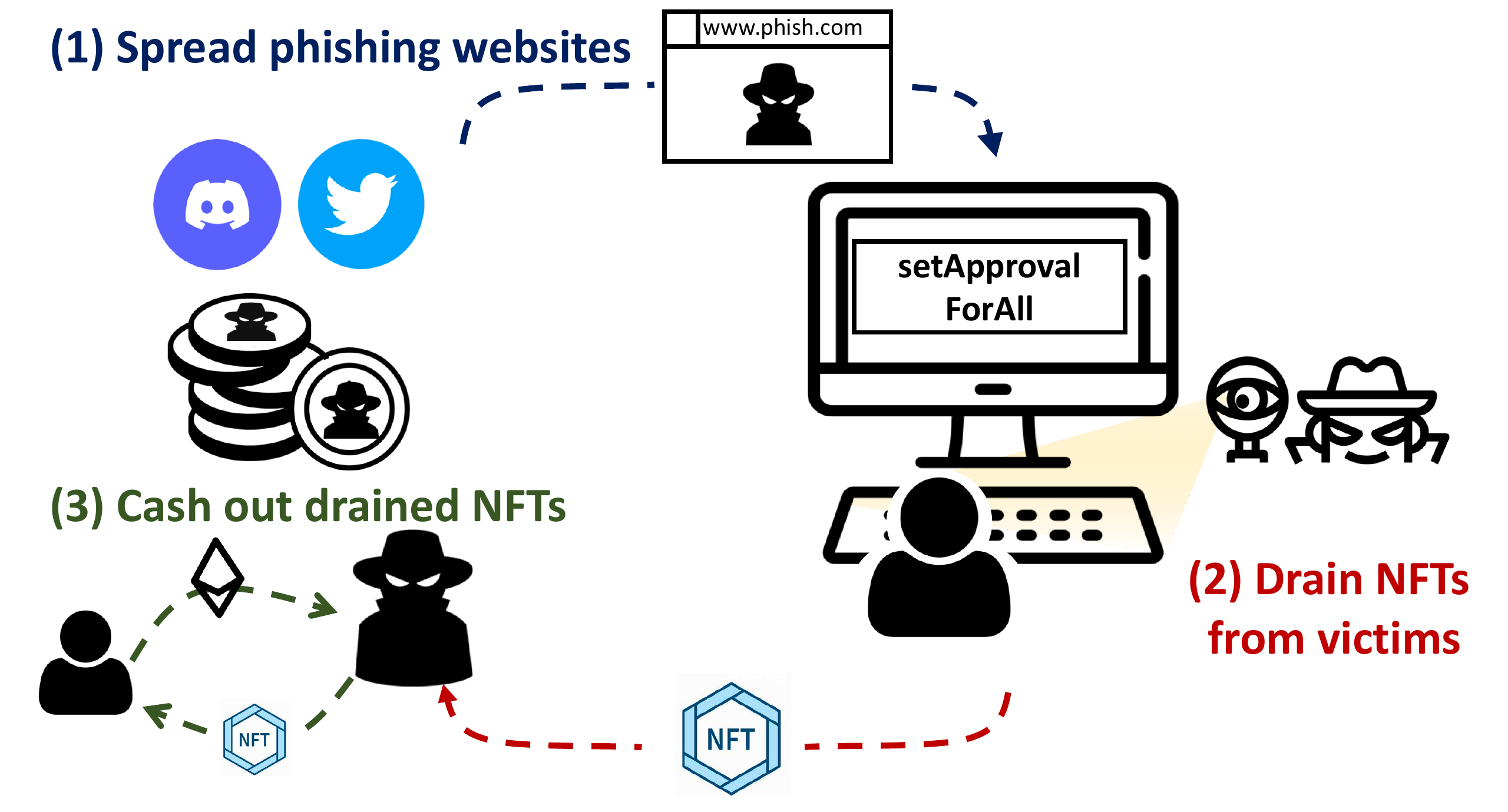}
    \caption{The process of draining NFTs using phishing attacks: (1) spreading phishing websites, (2) draining NFTs from victims, and (3) cashing out drained NFTs.}
    \label{f:drain_process}
    \vspace{-0.4cm}
\end{figure}

% \noindent\textbf{Phishing attacks in the NFT ecosystem} is commonly called \textbf{draining}~\cite{drain1, drain2}, and we will use this term in this paper.
\noindent\textbf{Stealing NFTs from victims' NFT wallets} is commonly called \textbf{draining}~\cite{drain1, drain2}, and we will use this term in this paper. 
The main goal of NFT drainers is to ``drain'' (steal) NFTs from victims, although cryptocurrencies and private information could also be targeted.
NFT drainers commonly use phishing scams for their purpose.
We detail the procedures of how NFT drainers operate by dividing them into three steps: (1) spreading phishing websites, (2) draining NFTs, and (3) cashing out drained NFTs. Figure \ref{f:drain_process} illustrates the process.

\textit{(1) Spread phishing websites.}
NFT drainers mainly use two methods to spread phishing websites to victims. 
First, they \textit{1.1) use social media}, such as Twitter. They can make social media accounts masquerading as official accounts of popular NFTs, sometimes even compromising them. The drainers upload posts linking to scam sites on these channels. Another method is to \textit{1.2) use phishing token airdrops}. Airdrops are commonly used in marketing campaigns to promote creators' projects by sending free tokens~\cite{airdrop}. However, it can also be abused by drainers to trick victims. A drainer can send fake tokens to target wallets, tricking victims into clicking a phishing website link in the token's description.

\textit{(2) Drain NFTs from victims.}
The drainers steal NFTs from lured victims \diff{through two primary methods}: \textit{2.1) \diff{capturing} login credentials of crypto wallets} or \textit{2.2) \diff{exploiting} an interface function}. 
The first method is similar to the traditional phishing methods that target victim passwords.
When victims \diff{input their crypto wallet credentials,} drainers can record the information using \diff{tools like} keyloggers~\cite{nft_scam}.
With the stolen credentials, drainers can access victim accounts\diff{to transfer NFTs into} their own wallets.
Alternatively, drainers can abuse smart contracts to deceive victims.
% Alternatively, drainers can abuse smart contracts to deceive victims who are not familiar with how they work. 
Drainers \diff{can craft} malicious smart contracts \diff{using functions} such as \textit{setApprovalForAll}, and lure victims to sign a transaction with the contract~\cite{setapproveforall}.
\textit{SetApprovalForAll} is an important method used for NFT trading. \diff{When selling NFTs on a marketplace, a user needs to call the function }to authorize the marketplace \diff{to transfer} the NFTs from the \diff{seller's} account to the buyer's.
However, by calling the function on a phishing website, a victim grants the drainer permission to transfer all the victim's tokens~\cite{token_approval}. 
\diff{On the blockchain, such an unauthorized transfer from the victim's account to the drainer's is logged as \textit{gifts}.}

%팔기
\textit{(3) Cash out drained NFTs.}
Scammers targeting cryptocurrencies cash out stolen assets using crypto exchanges and sometimes through mixing services~\cite{zhang2021blockchain, nft_encash}. Conversely, NFT drainers must sell the stolen NFTs by listing them on marketplaces~\cite{nft_encash}. 
To combat NFT scams, OpenSea~\cite{opensea}, the largest NFT marketplace, made a policy of disabling the buying, selling, and gifting of stolen items~\cite{opensea_policy}.\\

% \begin{definition}[\textbf{Transaction}] 
% There are four types of transactions in NFT space: mint, sale, transfer, and burn.
% \end{definition}

% \begin{definition}[\textbf{User Transaction}] 
% Sale and transfer are occurred between users, and we define them as user transaction. 
% Depending on in/out directions, they are divided by buy/sell and receive/send, respectively.
% \end{definition}
% \begin{figure*}[t]
%     \centering
%     \includegraphics[width=0.8\textwidth]{Fig/overview.png}
%     \caption{OverviewCaption}
%     \label{fig:overview}
% \end{figure*}

\subsection{Motivation}
NFTs have attracted the attention of investors and criminals alike. 
NFT drainers using phishing scams continue to make headlines~\cite{scam1, scam2, scam3}.
% Phishing scams targeting NFT users continue to make headlines~\cite{scam1, scam2, scam3}. 
For instance, a sophisticated phishing attack on Uniswap NFT holders caused damages of \$8 million when users were tricked into approving malicious transactions~\cite{uniswap1, uniswap2}.
% In another instance, the official Instagram account of the most valuable NFT, Bored Ape Yacht Club, was compromised and used in a phishing attack in April~\cite{bayc_hack1, bayc_hack2}. The drainers stole NFTs by luring victims with free tokens through the Instagram account. The total damage was valued at \$2.7 million, including a token worth \$354,500 (at that time of exchange rate). 

Alarmingly, there have been a wealth of tools made available to assist NFT draining. Software packages for NFT draining are being sold between \$29 – \$149 on dedicated websites~\cite{scam_as_service} and the darkweb~\cite{drain_darkweb}. 
% Some groups even publish a free version of their draining code on GitHub in order to advertise their online stores~\cite{scam_as_service}.
The accessibility of draining tools lowers the barriers to entry for future NFT drainers.
To prevent \diff{such} tools, MetaMask, a popular crypto wallet, updated to include a feature to warn users when transactions request the \textit{setApprovalForAll}~\cite{metamask1, metamask2} \diff{function}. However, NFT drainers have also developed tools to bypass this update~\cite{drainer_update}.

Although the damage caused by NFT drainers is increasing, no previous work has yet conducted an in-depth measurement study or proposed a detection method for NFT draining.
Several methods have been proposed to detect phishers targeting cryptocurrency~\cite{li2022ttagn, chen2020phishing, EGCN}, but they are unsuitable for applying to the NFT ecosystem for the following reasons.

First, essential features to detect cryptocurrency phishers do not apply to NFT drainers.
% For instance, NFT drainers do not follow the liquidation process of cryptocurrency phishers.
% Stolen cryptocurrencies are generally cashed out through exchanges after a mixing process, which is one of the most decisive features in detecting cryptocurrency phishers~\cite{chen2020phishing}.
% On the other hand, stolen NFTs cannot be laundered because each NFT is trackable. Also, NFTs cannot be directly cashed out in exchanges. Therefore, NFT drainers must list NFTs on the market for sale in order to cash out.
The ineffectiveness of previous features raises the need for a measurement study on NFT drainers to gain insights into how to detect phished NFTs.
In Section \ref{sec:measurement}, we investigate the activity of NFT drainers to identify trading traits that can detect NFT draining.

Second, existing methods cannot consider complex contexts within the NFT ecosystem.
Unlike cryptocurrency, where coins are interchangeable and have equal value, NFTs are of distinct identities with dynamic values, making complex transaction patterns (factoring price and frequency).
To fully understand a user's trading habits, the transaction history of each NFT \diff{the user trades should be considered.} 
In Section \ref{s:tx_context}, we describe our NFT drainer detection system, \model{}, which includes a module specifically designed to leverage a user's NFT transaction context.

Third, previous studies on detecting cryptocurrency phishing cannot fully capture the relationship between NFT users.
While cryptocurrency transactions are only limited to transfers, NFT transactions can be further classified into four categories.
% Particularly, there are two transaction types between NFT users (Figure \ref{f:nft_txtype}): \textit{sales} and \textit{gifts}. Our investigation shows that this distinction can be a significant indicator when interpreting relationships between users. 
Our investigation shows that \diff{the distinction of transaction types between NFT users is} a significant indicator when interpreting relationships between users. 
Section \ref{s:social_context} explains how \model{} uses a module to capture user relationships in the NFT trading network.

\section{Dataset Construction}
\label{sec:data}

This section summarizes our data collection approach and the datasets used in this study.
We collect two datasets in our paper: NFT transaction data and NFT drainer accounts. 

% \subsection{NFT transaction data}

\noindent\textbf{Fetching NFT transaction records from Ethereum blockchain:} 
There are two types of addresses in the Ethereum blockchain: Externally Owned Accounts (EOA) and Smart Contracts.
An EOA represents an account controlled by an individual.
It can send transactions, interact with smart contracts, and manage digital assets on Ethereum.
In this paper, we refer to EOAs simply as \textit{users} or \textit{accounts} to enhance readability. 
A smart contract is a code deployed on the blockchain executing actions and transactions based on predefined conditions.

\diff{We utilize \textit{transfer logs} to get a token's transmission history. Transfer logs are written to the blockchain by smart contracts whenever token transfers occur.}  With this information, we can identify changes in ownership of tokens, such as who sends the tokens to whom. 

% Token transfers, including those of NFTs and Fungible Tokens (FTs), are not recorded on the blockchain because they are internal transactions initiated by smart contracts. Thus, we use \textit{transfer logs}, which are written by smart contracts on the blockchain when token transfers occur, to understand a token's transmission history. With this information, we can identify changes in ownership of tokens, such as who sends the tokens to whom. 

We distinguish between two types of NFT transactions between users: sales and gifts.
However, because the NFT transfer logs do not have payment information, evidence of payment must be found from other sources.
\diff{\diffnote{MR3}During our research period, over 90\% of the NFT trading volume took place on three platforms: Opensea, Blur, and X2Y2~\cite{tx_trend}. A detailed analysis of the smart contracts used for NFT trading from these markets suggests that most NFTs traded on the marketplace are sold for Ether or various Fungible Tokens (FTs).
If an NFT buyer pays in Ether, it is recorded through an external transaction. 
If the buyer pays in FT, it is recorded in an FT transfer log. 
Hence, we gather both external transactions and transfer logs associated with both NFTs and FTs.}
To accomplish this, we run an Ethereum node and retrieve the data from the node using web3.py, an Ethereum-specific library to interact with the Ethereum blockchain.

\begin{figure}
\ifdiff
\begin{mdframed}[backgroundcolor=blue!10]
\fi
    \includegraphics[width=0.9\columnwidth]{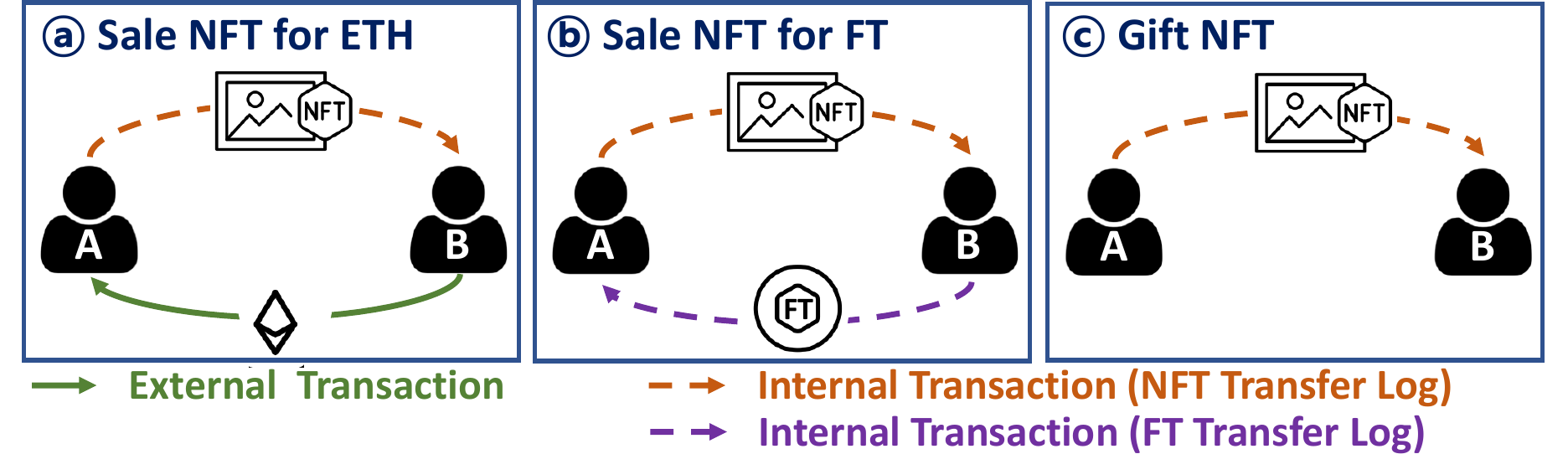}
    \caption{\diff{Visualizations simplifying the NFT transaction types on the Ethereum blockchain.}  
    We classify each NFT transaction record between users into sales (\textcircled{\small a} and \textcircled{\small b}) or gifts (\textcircled{\small c}) based on whether a payment was made.}
    \label{fig:NFT_collect}
    \vspace{-0.4cm}
\ifdiff
\end{mdframed}
\fi
\end{figure}

\diff{Figure \ref{fig:NFT_collect} provides an overview of how NFT transaction types appear on the Ethereum blockchain.
If there is an NFT transfer log from user $A$ to user $B$ and an external transaction in which $B$ sends Ether to $A$, we infer the NFT was sold for Ether (\textcircled{\small a}).
Instead, if there is an FT transfer log in which $B$ sends FTs to $A$, we interpret this as the NFT being sold for FTs (\textcircled{\small b}). }
\diffnote{MR3}\diff{For both sale types, we include indirect sales, such as when an NFT sender receives currency from the NFT receiver through a marketplace address.  
Otherwise, we regard that there was no payment for the NFT, and the NFT's transfer from $A$ to $B$ is considered as a gift (\textcircled{\small c}).

\diffnote{MR3} To confirm the quality of our NFT transaction data, we analyzed weekly transaction statistics of our data, which include the number of sales, the number of gifts, and trading volume.
Our data collection is supported by similar reportings from NFTGo~\cite{nftgo_data}, a renowned NFT analysis platform.

}

\begin{table}[t]
\caption{Summary of collected NFT transaction records. Note that the number of mintings is larger than the number of NFTs due to the multiple mintings of spam NFTs.}
\label{tab:NFT_tx}
\centering
\footnotesize
\ra{1.03}
% \resizebox{\linewidth}{!}{
\begin{tabular}{l l r}
    \toprule
    \multicolumn{2}{l}{\textbf{Type}} & \textbf{Collection} \\
    \midrule
    \multicolumn{2}{l}{NFT} & 80,795,833 \\
    \midrule
    \multicolumn{2}{l}{Address} & 4,733,670\\
    & EOA (Account) & 4,640,645 \\
    & Smart contract & 93,025\\
    \midrule
    \multicolumn{2}{l}{Transaction} & 127,820,930\\
    & Mint & 81,769,127 \\
    &  Sale & 24,915,481  \\
    &  Gift & 19,722,551\\
    &  Burn & 1,413,771\\ 
    \midrule
    \midrule
    \multicolumn{2}{l}{\textbf{Period}} &
    \multicolumn{1}{l}{January 1,2022 $\sim$ December 31, 2022}\\
    \bottomrule
\end{tabular}
%}
\vspace{-0.2cm}
\end{table}

\label{s:NFT_transaction}

\begin{figure*}[t]
     \centering
    \includegraphics[width=0.95\textwidth]{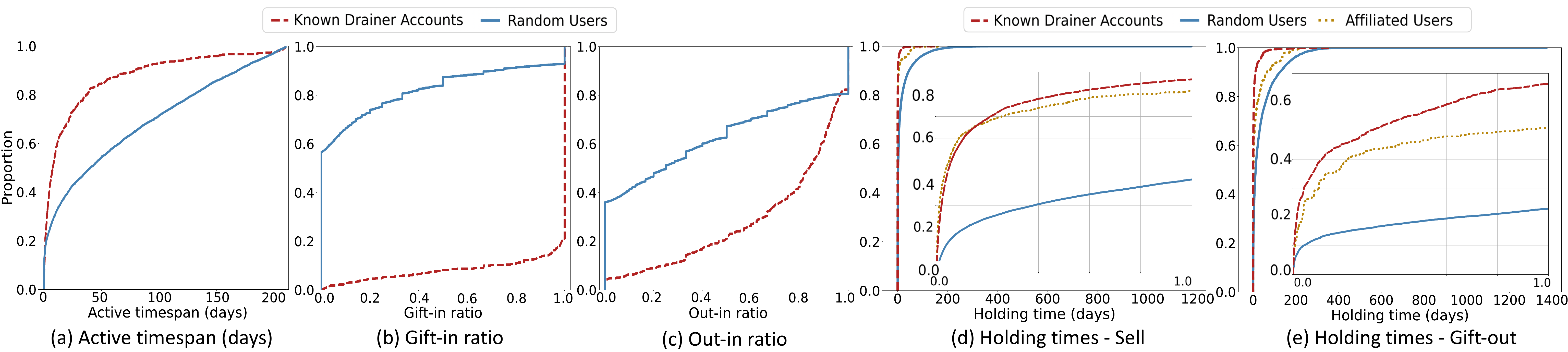}
        \caption{CDFs of (a) active timespan, (b) gift-in ratio, and (c) out-in ratio, where the y-axis denotes proportion of users.
        CDFs of holding times of different user types (regular users, affiliated users, and drainers) based on out-transaction types: (d) \textit{sell} and (e) \textit{gift-out}, where the y-axis denotes proportion of NFTs. We include an enlarged graph for holding times smaller than a day. }
        \label{f:behave_pattern}
        \vspace{-0.3cm}
\end{figure*}

According to Elliptic\cite{elliptic}, NFT draining grew rapidly and has been on the rise since January 2022~\cite{elliptic_ana}. As a result, we focus on data with \texttt{block\_timestamp} between January 1, 2022 to December 31, 2022.
We obtain more than 127 million transactions from this period for 80 million NFTs. 
The data includes transactions from over 4 million unique accounts.
Our NFT transaction data is summarized in Table \ref{tab:NFT_tx}. 
% In our analysis, we only consider \textit{mints}, \textit{sales}, and \textit{gifts} because transaction records of \textit{burns} account for less than 1\% of total transactions.

\begin{table}[t]
\caption{Summary of collected drainer accounts from various channels.}
\label{tab:collect_label_data}
\centering
\ra{1.03}
\footnotesize
\begin{tabular}{l r r}
    \toprule
    \textbf{Channel} & \textbf{\# active accounts} 
    & \textbf{\# drainer accounts}\\
    \midrule
    Scamsniffer & 817 & 797 \\ 
    Chainabuse & 769 & 737 \\
    Twitter	&728	&682 \\
    Etherscan	&128	&128 \\
    CryptoscamDB	&68	&66 \\
    \midrule
    Total &	1,230 & 1,135 \\
    \midrule
    \midrule
    \multicolumn{1}{l}{\textbf{Period}} &
    \multicolumn{2}{r}{January 1,2022 $\sim$ December 31, 2022}\\
    \bottomrule
\end{tabular}
\vspace{-0.2cm}
\end{table}

% \begin{table}[t]
% \caption{Summary of collected drainer accounts from various channels.}
% \label{tab:collect_label_data}
% \centering
% \ra{1.1}
% \footnotesize
% % \resizebox{\linewidth}{!}{
% \begin{tabular}{lr lr}
%     \toprule
%     \textbf{Channel} & \textbf{\# accounts} & \textbf{Channel} & \textbf{\# accounts}\\
%     \midrule
%      ScamSniffer & 462 & Chainabuse & 150 \\
%      Etherscan & 127 & Twitter & 105 \\
%     CryptoScamDB & 49 \\
%     \midrule
%     \multicolumn{2}{l}{Total} & \multicolumn{2}{r}{742} \\
%     \midrule
%     \midrule
%     \multicolumn{2}{l}{\textbf{Period}} &
%     \multicolumn{2}{r}{January 1,2022 $\sim$ August 31, 2022}\\
%     \bottomrule
% \end{tabular}
% % }
% \end{table}

% \begin{table}[t]
% \caption{Summary of collected drainer accounts from various channels.}
% \label{tab:collect_label_data}
% \vspace{-0.2cm}
% \centering
% \ra{1.1}
% \footnotesize
% % \resizebox{\linewidth}{!}{
% \begin{tabular}{lr lr}
%     \toprule
%     \textbf{Channel} & \textbf{\# accounts} & \textbf{Channel} & \textbf{\# accounts}\\
%     \midrule
%      ScamSniffer & 462 & Chainabuse & 150 \\
%      Etherscan & 127 & Twitter & 105 \\
%     CryptoScamDB & 49 \\
%     \midrule
%     \multicolumn{2}{l}{Total} & \multicolumn{2}{r}{742} \\
%     \midrule
%     \midrule
%     \multicolumn{2}{l}{\textbf{Period}} &
%     \multicolumn{2}{r}{January 1,2022 $\sim$ August 31, 2022}\\
%     \bottomrule
% \end{tabular}
% % }
% \vspace{-0.3cm}
% \end{table}

\noindent\textbf{Crawling NFT drainer accounts:}
We crawled drainer accounts reported for NFT phishing scams from five websites: Twitter, ScamSniffer, Etherscan, CryptoScamDB, and Chainabuse.

(1) Twitter~\cite{twitter} is utilized by NFT users as a channel to share information on drainer accounts. We first collect tweets that mention the phrase ``NFT'' and one of the following keywords: \textit{phish, hack, drain, stole}. 
\diffnote{MR3}\diff{Then, we manually select tweets written by high-profile users who analyze scammers in the crypto-space professionally.}
(2) ScamSniffer~\cite{scamsniffer} collects malicious accounts by visiting suspicious sites that trick users into making dangerous transactions.
\diff{ScamSniffer checks whether accounts have suspicious behavior through other security services~\cite{scamsniffer_data}.}
(3) Etherscan~\cite{etherscan}, an Ethereum block explorer, provides a list of Ethereum addresses reported for phishing/hacking.
\diff{Etherscan reviews and assesses reports to prove the accounts are involved in scams or phishing activities~\cite{etherscan_data}.}
% They receive reports  
(4) CryptoScamDB~\cite{cryptoscamdb} collects reports on scams in the crypto space, including malicious accounts, URLs, and descriptions.
\diff{CryptoscamDB manually scans the reports before adding addresses to the dataset~\cite{cryptoscamdb_data}.
We only select reports related to NFT draining by using the same criteria in (1) Twitter.} 
(5) Chainabuse~\cite{chainabuse} provides scam reports with descriptions across multiple blockchains. 
\diff{Chainabuse has a spam detection system and attributes a confidence score to reports calculated by experts~\cite{chainabuse_1}. We collected Ethereum addresses with the `Checked by Chainabuse' badge~\cite{chainabuse_2} reported under the NFT Scam category. }

% \diffnote{MR3}\diff{We believe that the quality of our collected drainer accounts is reliable. Firstly, the reporting services have a system to verify reported drainer accounts. 
% For instance, ScamSniffer checks whether accounts have suspicious behavior through other security services~\cite{scamsniffer_data}. 
% Chainabuse has a spam detection system and attributes a confidence score to reports calculated by experts~\cite{chainabuse_1}. We collected reports with the 'Checked by Chainabuse' badge~\cite{chainabuse_2}. 
% CryptoscamDB manually scans the raw reports before adding addresses to the dataset~\cite{cryptoscamdb_data}.
% For Twitter data, we manually selected tweets written by high-profile users who analyze scammers in the crypto-space professionally.}

We gathered reports from the above channels based on \diff{the specified criteria until} January 1, 2023. 
Note that \diff{some accounts were reported multiple times across} different sites.
During \diff{our} data collection period, \diff{we identified} 1,230 accounts involved in NFT transactions.
\diffnote{R2C1}\diff{Since not all reported accounts were successful, some reported accounts did not show drainer activity.}
As \diff{outlined} in Section~\ref{s:background}, the act of draining NFTs from victims' wallets is categorized as \textit{gifts}. 
\diff{Based on this, we define accounts that have at least one gifted-in NFTs as \textit{drainers}.} Using this definition, we identified 1,135 unique accounts labeled as drainers (summarized in Table \ref{tab:collect_label_data}).

% Among the collected labeled accounts,

\section{Drainer Activity Characterization}
\label{sec:measurement}
Since drainers have malicious intent, they are likely to exhibit different behavior patterns in NFT trading from regular users. 
In this section, we look into distinguishable traits of drainers that motivate the design of our NFT drainer detection model. 
We conduct a measurement study with NFT transaction records from January 1, 2022, to July 31, 2022, including 645 drainer accounts.
First, by comparing primary trading features with regular users, we verify that most drainer accounts are used only for draining.
Then, we focus on their liquidation patterns and how they are distinct from regular transactions. 
We track drained NFTs and find that they have different \textbf{NFT transaction context} and \textbf{social context} from regular users. 
% that they use alternate accounts, sell at a fast speed, and sell at bargain prices.
% \subsection{NFT drainer accounts}

%\subsection{Basic trading of drainers}
\subsection{Trading behavior of drainers}
\label{s:basic}
As mentioned in \ref{s:background}, there are many efforts to prevent phishing scams in the NFT space, such as sharing information about drainers, phishing websites, and drained NFTs. 
If drainer accounts become known to be suspicious, it is nearly impossible for them to trade NFTs with benign users. Thus, we expect that drainers are more likely to 1) have short-lived accounts and 2) use their accounts only for draining. 

In order to investigate these two assumptions, we analyze the trading activity of 645 NFT drainers. 
For comparison, we randomly extract 10,000 users from our data of 3 million regular users.
From this set, users with only one transaction record were excluded since their active timespan cannot be calculated. This left 6,658 regular users to be used in the analysis.
We compare these two groups along three dimensions. 

First, we measure how long they traded by calculating the active timespan: the time difference (in days) between the first and last transaction recorded for each user. 
Figure ~\ref{f:behave_pattern}(a) shows that drainers trade for a shorter period than regular users.
This phenomenon is also similarly observed in Ether phishing accounts~\cite{chen2020phishing}.

Second, we look into how NFT drainers utilize their accounts. Previous work to detect cryptocurrency phishers~\cite{li2022ttagn,chen2020phishing} only focused on the amount and frequency of user transactions.
However, multiple transaction types exist in NFT trading. This is an important consideration since we assume drainers will be gifted NFTs from victims (as opposed to buying or minting NFTs).
Thus, we calculate the ratio of gifting-in to all in-transaction types (gifting-in, buying, and minting). 
In Figure~\ref{f:behave_pattern}(b), the majority of drainers have a relatively high proportion of NFTs gifted; 75.1\% of drainers obtain NFTs only through gifting-in in the three in-transaction types. 

In addition, we analyze how likely drainers are to transfer out.
Drainers looking to liquidate do not wish to hold their NFTs. Therefore, drainers are likely to transfer their NFTs out. We measure \textit{out-in ratio}, the ratio of the number of out-transactions to in-transactions. 
% Note that the in-out ratio can be greater than 1, because users could sell or send their NFTs obtained before the time our data covers. For better visualization, we transform the number greater than 1 to 1. 
In Figure~\ref{f:behave_pattern}(c), we observe drainers have a higher out-in ratio than regular users. 
Regular users generally have a lower out-in ratio: 38.1\% of them did not make any out-transactions at all. On the other hand, drainers have higher out-in ratios: 75.9\% of them make out transactions on more than half of their NFTs. This suggests drainers have different intentions from regular users, who are more likely to use NFTs for collecting or investing.

By combining the observations above, we conclude that most drainer accounts are for draining purposes only. Additionally, we further analyze differences between drainers and regular users along 19 dimensions (more details in Appendix~\ref{apx:feat_analysis}) and utilize them in our detection method.

%drainer가 victim한테 NFT를 훔치는 것은 NFT가 victim으로 부터 drainer 지갑으로 transfer되는것과 같음. 즉, drainer의 input 중 receive 비율이 높을 것임. 해봤더니, 80%가 !! --> 대부분 drainer용으로 사용된 듯또한, receive 속도가 빠름 --> 이거는 위에서 말한 신고 당하면 활동 못해서 

% \begin{table}[t]
%     \caption{Statistics on transactions until drained NFTs are sold. 훔쳐진 NFT와 그 중 팔린 애들 }
%     \label{tab:after_drain}
%     \centering
%     \resizebox{0.9\linewidth}{!}{
%         \begin{tabular}{l rr rr r}
%             \toprule
%             \multicolumn{1}{l}{} && \multicolumn{2}{l}{\textbf{Total}} & \multicolumn{2}{l}{\textbf{Sold}} \\
%             \cmidrule(rl){3-4} \cmidrule(rl){5-6}
%             \multicolumn{1}{l}{} & \textbf{\# transfer} & \textbf{\#} & \textbf{\%} & \textbf{\#} & \textbf{\%} \\
%             \midrule
%             Sale & 0& 11,195 & 41.8 & 11,195 & 100 \\
%             \midrule
%             \multirow{2}{*}{Transfer} & 1 & 6,426 & 24 & 4,825 & 75.1 \\
%               & $\geq$ 2 & 1,125 & 4.2 & 442 & 39.3 \\
%             \midrule
%             None & 0 & 8,065 & 30.1  & 0  & 0 \\
%             \midrule
%             \multicolumn{2}{c}{Total}             & 26,811        &            & 16,462        &          \\
%             \bottomrule
%         \end{tabular}
%     }
% \end{table}

\begin{table}[t]
    \caption{Statistics on transactions about drained NFTs according to behaviors right after draining: \textit{sell}, \textit{gift-out}, and \textit{none}(just holding).
    \# gifting is the number of giftings before the first sale after draining, and \% sold is the percentage of NFTs eventually sold. }
    \vspace{-0.1cm}
    \label{tab:after_drain}
    \centering
    \ra{1.1}
    \footnotesize
        \begin{tabular}{l r r r}
            \toprule
            \multicolumn{1}{l}{\textbf{Type}} & \textbf{\# gifting} &
            \textbf{\# drained NFTs} & \textbf{\% sold} \\
            \midrule
            Sell & 0 & 11,195 (41.8\%) & 100\% \\
            \multirow{2}{*}{Gift-out} 
            & 1 & 6,426 (24.0\%) & 75.1\% \\
            
            & $\geq$ 2 & 1,125 (4.20\%) & 39.3\% \\
            None & 0 & 8,065 (30.1\%)  & 0\% \\
            \midrule
            \multicolumn{2}{l}{Total}  & 26,811 (100\%) &            61.4\% \\
            \bottomrule
        \end{tabular}
        \vspace{-0.4cm}
\end{table}

\subsection{Liquidation behavior of drainers}
\label{s:liquidation}
In this section, we dive deep into the liquidation process of drained NFTs. 
The uniqueness of NFTs enables us to track the transaction history of each NFT.
Since drainer accounts are only used for draining, we assume all gifted-in NFTs were stolen from victims.
% From the premise,  drainers' accounts are used for draining only, made in Section \ref{s:basic}, we regard the NFTs received by drainer accounts as drained from victims. 

% Further, 72.7\% of transferred NFTs are sold from the recipients. 
%39명이 한번도 out X, 52명 : send만, 113명 : sell만 

 \noindent\textbf{Alternate accounts.}
We find a total of 26,811 NFTs that were gifted to drainers (Table \ref{tab:after_drain}). 
Of these, 41.8\% were sold directly, 28.2\% were gifted-out to other users, and the rest (30.1\%) remained in the drainer's wallets. 
\diffnote{R1C1}\diff{We defined \textit{affiliated users} as all accounts that were gifted drained NFTs from known drainer accounts. }

% Transferring NFTs without paying between users usually implies that they have in a close tie in NFT ecosystem~\
% \textit{gifting} NFTs between two users implies that they are in a specific relationship. 

Only 17.5\% of drainers do not have affiliated users.
In other words, most drainers (82.5\%) have one or more affiliated users and use them to liquidate drained NFTs indirectly.
We find 637 affiliated users including 15.4\% that are related to two or more drainers. The most overlapped affiliated user is connected with 16 drainers.
% (more details are in Appendix~\ref{apx:connection}). 
The fact that many drainers choose to liquidate through the same affiliated accounts suggests a close relationship between them.
60\% of affiliated users get NFTs only through gifting-in, while the rest of them (40\%) participate in buying and minting like regular users (Appendix~\ref{apx:feat_analysis_1}).
From the result, we observe that most affiliated users exhibit similar behaviors to drainer accounts, but there are also a significant number of affiliated accounts that engage in general NFT trading.

\noindent\textbf{Rapid liquidation:} We now analyze how quickly drainers liquidate NFTs by measuring the holding times of each NFT. We define holding time as the timespan a user held ownership of an NFT, and measure it by taking the difference (in days) of the in-transaction and out-transaction. 
Holding times can vary greatly depending on the user's investment strategy and characteristics of the NFT, such as market price and rarity. 
We measure the holding times of all users that owned the NFT for each of the 18,746 drained NFTs with out-transactions.
Note that the holding time can be longer than seven months (our collection period) because we refer to NFT transaction data before 2022 to minimize the bias.
We compare drainer holding times with those of affiliated users and those of regular users along two out-transaction types: sell and gift-out. 

%drainer가 판다는 것은 자기가 직접 파는거, asso가 파는 것은 drainer가 asso한테 보내서 v
Drainers and affiliated users show similar distributions which are noticeably different from that of regular users.
Figure~\ref{f:behave_pattern} (d) shows they sell more than 80\% of NFTs within a day, which is twice more likely than regular users. They also have short holding times before gifting, but drainers gift NFTs much faster compared to affiliated users (Figure~\ref{f:behave_pattern} (e)).
By integrating these observations with the fact that up to 75\% of NFTs sent to affiliated users are sold (Table \ref{tab:after_drain}), we notice that drainers alternate between two strategies of liquidation: (1) directly selling NFTs quickly or (2) quickly gifting-out NFTs to affiliated users for selling.

\diff{For comparison with regular transactions,} we calculate the average holding time ($HT\_regular_{avg}$) for each NFT, \diff{which serves as} a reference to a drainer’s holding time ($HT\_drain$). 
\diffnote{R2C2} \diff{Note that the ($HT\_regular_{avg}$) is calculated from the NFT’s minting date up to July 2022.}
As shown in Table \ref{tab:ht_cmr}, 90\% of $HT\_drain$s are shorter than $HT\_regular_{avg}$s, with an average of 87.7\% percent decrease regardless of the out-transaction types. We also observe that $HT\_drain$ is the minimum holding time for 70\% of NFTs. From these results, we identify that drainers deviate from regular transaction patterns in terms of holding time.

\begin{table}[t]
    \caption{Statistics of holding times of drained NFTs depending on user type along out-transaction types: \textit{sell} and \textit{gift-out}. 
    % $HT\_regular_{avg}$ ($HT\_regular_{min}$): average (minimum) of regular users' holding time. $HT\_drain$: drainers' holding time. 
    Percent decrease is the measure of the decrease from $HT\_regular_{avg}$ to $HT\_drain$ as a percentage of $HT\_regular_{avg}$.
    }    
    \label{tab:ht_cmr}
    \centering
    \footnotesize
        \begin{tabular}{l r r}
            \toprule
            & \multicolumn{2}{r}{\textbf{\# drained NFTs }}\\
            \cmidrule{2-3}
             \textbf{Case} & \textbf{Sell} & \textbf{Gift-out} \\
            \midrule
            $HT\_drain$ $<$ $HT\_regular_{avg}$
            & 8,077 (88.9\%) & 6,210 (90.9\%) \\ 
            $HT\_drain$ =  $HT\_regular_{min}$
            & 6,498 (71.4\%) & 5,034 (73.7\%)  \\
            \midrule
            Total & 9,085 (100\%) & 6,832 (100\%) \\
            \midrule
            \midrule
            \multicolumn{3}{l}{\textbf{Stats of percent decrease}}\\
            \midrule
            \multicolumn{1}{l}{Mean} & 87.7\% & 87.5\% \\
            \multicolumn{1}{l}{Standard deviation} & 29.5 & 29.8 \\
            \bottomrule
        \end{tabular}
\end{table}

% \begin{table}[h]
%     \caption{Statistics on NFT holding time depending on user type.  HT\_regular\_\textit{avg} (\textit{min}): average (minimum) of regular users' holding time. HT\_drain: drainers' holding time.}
%     \label{tab:ht_cmr}
%     \centering
%     \resizebox{\linewidth}{!}{
%         \begin{tabular}{ccccc}
%             \toprule
%             \multicolumn{1}{c}{} & \multicolumn{2}{c}{\textbf{Sell}} & \multicolumn{2}{c}{\textbf{Send}} \\
%             \cmidrule(rl){2-3} \cmidrule(rl){4-5} 
%             \textbf{Case} & {\#} & {\%} & {\#} & {\%} \\
%             \midrule
%             HT\_regular\_\textit{avg} > HT\_drain 
%             & 8,077 & 88.9\% & 6,210 & 90.9\% \\ 
%             HT\_regular\_\textit{min} = HT\_drain 
%             & 6,498 & 71.4\% & 5,034 & 73.4\%  \\
%             \midrule
%             \multicolumn{5}{c}{\textbf{Stats of the decrease rate in HT\_drain compared to HT\_regular}}\\
%             \midrule
%             \multicolumn{1}{c}{average} & \multicolumn{2}{c}{87.7} & \multicolumn{2}{c}{87.5} \\
%             \multicolumn{1}{c}{standard deviation} & \multicolumn{2}{c}{29.5} & \multicolumn{2}{c}{29.8} \\
%             \bottomrule
%         \end{tabular}
%     }
% \end{table}

\begin{table}[t]
    \caption{Statistics on sale price of drained NFTs compared to their market price ($\text{Price}_{avg}$, $\text{Price}_{closest}$). 
    Only consider NFTs sold after draining with other sale records of more than one. 
    % $\text{Price}_{avg}$: the average sale price of each NFT.
    % $\text{Price}_{closest}$: the closest sale price of each NFT from draining.
    % $\text{Price}_{drain}$: sale price of drained NFT by drainers/affiliated users.
    \textit{p.d.} denotes the measure of decrease from each market price to $\text{Price}_{drain}$ as a percentage of the market price.}
    \label{tab:price_cmp}
    \centering
    \footnotesize
        \begin{tabular}{lrrr}
            \toprule
            \multicolumn{1}{c}{} & & \multicolumn{2}{r}{\textbf{Stats of p.d.}} \\
            \cmidrule(rl){3-4}
            \textbf{Case} & \multicolumn{1}{r}{\textbf{\# drained NFTs}}  & {Mean} & {Std} \\
            \midrule
            $\text{Price}_{avg}$ $>$ $\text{Price}_{drain}$
            & 8,214 (74.0\%) & 37.3\% & 24.9\% \\ 
            $\text{Price}_{closest}$ $>$ $\text{Price}_{drain}$ 
            & 8,490 (76.5\%) & 39.1\% & 24.2\%  \\
            \midrule
            Total  & 11,100 (100\%) && \\
            \bottomrule
        \end{tabular}
\end{table}

% \begin{table}[t]
%     \caption{Statistics on NFT price depending on user type. We only consider NFTs that sold after draining and have other sale records. *dr: The decrease rate of NFTs sold cheaper. }
%     \label{tab:price_cmp}
%     \centering
%     \resizebox{\linewidth}{!}{
%         \begin{tabular}{cccc}
%             \toprule
%             \multicolumn{1}{c}{} & & \multicolumn{2}{c}{\textbf{Stats of dr*}} \\
%             \cmidrule(rl){3-4}
%             \textbf{Case} & \multicolumn{1}{c}{\textbf{\# NFTs}}  & {average} & {standard deviation} \\
%             \midrule
%             $\text{Price}_{average}$ > $\text{Price}_{drain}$
%             & 8,214 (74.0\%) & 37.3 & 24.9 \\ 
%             $\text{Price}_{closest}$ > $\text{Price}_{drain}$ 
%             & 8,490 (76.5\%) & 39.1 & 24.2  \\
%             \bottomrule
%         \end{tabular}
%     }
% \end{table}

\noindent\textbf{Bargain prices:} \diff{The findings from our measurements naturally} raise a question; \textit{how \diff{can} drainers liquidate drained NFTs so quickly?} To answer this question, we compare the sales prices of drainers/affiliated users ($\text{Price}_{drain}$) with the market prices. 
However, unlike stocks or cryptocurrencies, each NFT has
its own unique value. 
\diff{Also, market prices are susceptible to fluctuations based on supply and demand dynamics~\cite{price_change}. }
\diff{Thus, defining a market price for an NFT is feasible only when a sale occurs.}

\diff{To instead provide a comparative market price, we employ two baselines: $\text{Price}_{avg}$ and $\text{Price}_{closest}$. $\text{Price}_{avg}$ \diffnote{R2C2} represents the mean sales price from the NFT’s minting date up to July 2022, and $\text{Price}_{closest}$ signifies the sales price from a transaction occurring nearest to the drainer's sale time.}
We observe that drainers sell 74\% and 76\% of their NFTs cheaper than the two baselines, respectively, with an average price decrease of 37\% and 39\% (Table \ref{tab:price_cmp}).

% To sum up, most drainers liquidate their NFTs quickly. They do this by selling them at a lower price than the market price. They sometimes make these sales through affiliated users. This fast liquidation behavior is consistent with the scenario in which drainers must sell stolen NFTs before they are frozen by marketplaces for suspicious activity. We verify that the process of drainers stealing and liquidating NFTs deviates from the regular NFT transaction patterns. 

In summary, our analysis \diff{reveals distinct transaction and social contexts exhibited by NFT drainers compared to regular users.} 
Specifically, drainers tend to have irregular \textbf{NFT transaction contexts}, such as quickly liquidating NFTs at prices lower than the market value. 
Additionally, drainers often have unique \textbf{social contexts}, such as making sales through affiliated users and \diff{often are linked to the same affiliated users.} 

\section{Design of NFT Drainer Detector}
\label{sec:design}

\begin{figure*}[t]
    \centering
    \includegraphics[width=0.8\textwidth]{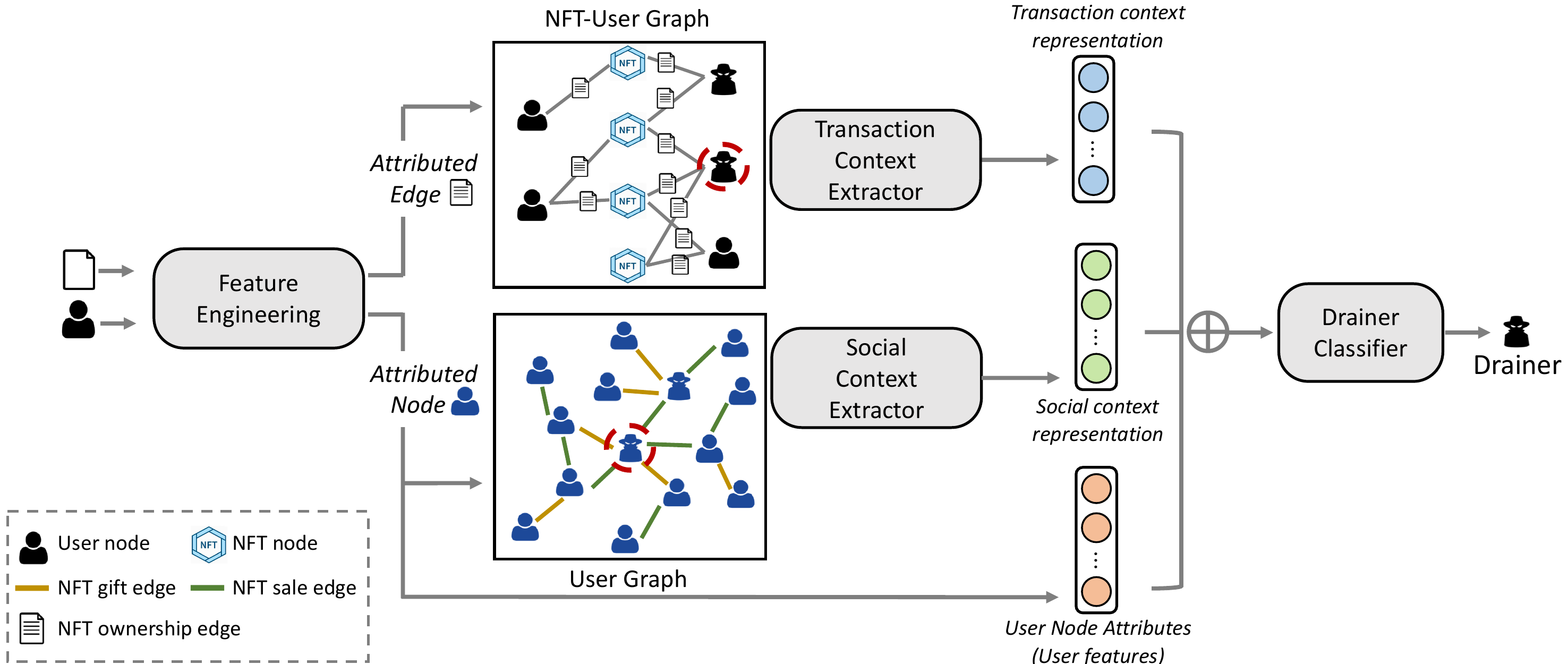}
    \caption{The overall architecture of \model{}. First, NFT transaction records and user accounts obtain attributes through feature engineering and are used to construct an NFT-User graph and a User graph. The transaction context extractor and social context extractor are trained to extract information from each graph, respectively. The user features and the representations from two extractors are concatenated and fed into the classifier.}
    \label{fig:overall}
    \vspace{-0.4cm}
\end{figure*}

\diff{Section~\ref{sec:measurement} emphasizes the importance of understanding both transaction and social contexts for NFT drainer detection.}
However, capturing the intricate relationships between millions of users
and NFTs for drainer detection remains \diff{a challenge}.

\diff{To tackle the challenge,} we designed a graph-based NFT drainer detection model, \model{}, depicted in Figure~\ref{fig:overall}.
The model creates a comprehensive representation of each user using a transaction context extractor and a social context extractor. 
Each extractor is trained to learn the relevant context on a \textit{NFT-User graph} and a \textit{User graph}, respectively. 
The NFT-User graph models NFT-to-user interactions using \textit{NFT ownership edge attributes}, and the user transaction contexts can be fully captured by aggregating the interaction history of all of their owned NFTs.
The User graph models comprehensive user-to-user interactions using \textit{user node attributes} \diff{that detail user trading behaviors and 
two types of edges representing user interactions.}
The two contexts are then combined \diff{with user node attributes} and fed into a classifier to integrate the information.
% First, we perform feature engineering to get two types of attributes, NFT ownership attributes and user attributes.
% NFT ownership attributes represent how users interact with NFTs in a \textit{User-NFT graph}, and user attributes represent users' detailed trading behavior considering NFT characteristics in the User Graph.
% Then, transaction context representations are extracted from the User-NFT Graph, and social context representations are extracted from the User Graph. 

% The drainer classifier creates a complex representation of each user, which comes from three parts: feature engineering, a transaction context extractor, and a social context extractor.  
% First, we perform feature engineering to get two types of attributes, NFT ownership attributes and user attributes.
% NFT ownership attributes represent how users interact with NFTs in a \textit{User-NFT graph}, and user attributes represent users' detailed trading behavior considering NFT characteristics in the User Graph.
% Then, transaction context representations are extracted from the User-NFT Graph, and social context representations are extracted from the User Graph. 

\subsection{Feature Engineering}
Before learning high-level user representations, we perform feature engineering based on observations in Section~\ref{sec:measurement} to obtain \textit{NFT ownership edge attributes} and \textit{user node attributes}.

\subsubsection{NFT ownership edge attributes}
\label{s:nft_tx_att}
% In Section \ref{sec:measurement}, we identify that drainers exhibit different liquidation behavior from regular users. 
To capture different transaction behaviors, we create representations of how users interact with NFTs, which is used in the NFT-User graph. 
We use the following 7 features to represent NFT ownership:

\begin{enumerate}
    \item \textbf{Holding time:} 
    Holding time is the timespan a user held ownership of the NFT. If there is no out-transaction, we calculate it by taking the difference (in days) between the in-transaction and the last day of our collection period.
    As drainers tend to sell or gift NFTs faster than regular users, their holding times are shorter than those of regular users. 
    \item \textbf{In-transaction type \& Out-transaction type:}
    Each transaction type is a categorical feature of how the user received the NFT (buy or gift-in) and what the user did with the NFT (sell, gift-out, or hold), respectively.
    Drained NFTs have the gift-in transaction type.
    \item \textbf{In-price \& Out-price:} 
    Each is the price (in Ether) the user sent to receive the NFT and the price the user received when sending the NFT out, respectively.
    It is zero if the transaction type is gifting, and -1 if no out-transaction was made.
    The in-price and out-price are significant because drainers sell their NFTs cheaper than regular transactions.
    \item \textbf{Average holding time \& Average sale price:} For comparison, we include the NFT's average holding time and sales price across all owners, \diff{calculated over the entire lifespan of the NFT.} \diffnote{R2C2} This can be used as a reference in finding anomalies in holding times and pricing. 
\end{enumerate}

\subsubsection{User node attributes}
\label{s:user_att}
To comprehensively understand the social relationships between users, we created detailed representations of their trading behavior, which is used in the User graph. We introduce 19-dimensional user node attributes that take into account NFT characteristics such as collections and transaction types, which helps the model to identify distinct behavior patterns of drainers.

\begin{enumerate}
    \item \textbf{Active timespan:} The time difference between the first and the last transaction. Drainers are more likely to have a short active timespan than regular users.
    \item \textbf{Gift-in ratio:} The ratio of gifting-in to all in-transactions (minting, buying, gifting-in). 
    Most drainers obtain NFTs only through gifting.
    \item \textbf{Out-in ratio:} The ratio of out-transactions to in-transactions.
    Drainers have a higher out-in ratio than regular users.
    \item \textbf{Number of each transaction type:} Transaction types considered are minting, buying, gifting-in, selling, and gifting-out. Drainers participate in selling and gifting-in more than other types. 
    (Appendix~\ref{apx:feat_analysis_1})
    \item \textbf{Number of collections for each transaction type:} NFT users commonly form communities based on collections~\cite{nadini2021mapping} and trade NFTs from similar collections. Unlike regular users, drainers tend to trade more kinds of collections than regular users. The transaction types are the same as above.
    (Appendix~\ref{apx:feat_analysis_2})
    \item \textbf{Number of neighbors for each transaction type:} Accounts that made a transaction with each other are considered neighbors. Drainers tend to have more neighbors from gift-in transactions. The transaction types are the same as above, excluding minting (which produces no neighbors). (Appendix~\ref{apx:feat_analysis_3})
    \item \textbf{Frequency of gift-ins \& sales:} The number of gifting-in/selling transactions divided by the active timespan. Gift-ins and sales occur more frequently for drainers. (Appendix~\ref{apx:feat_analysis_4})

\end{enumerate}

\subsection{NFT Transaction Context Extractor}
\label{s:tx_context}
% Drainers have distinguishable NFT transaction patterns compared to regular users. 
% However, trading traits are complex to represent because each NFT has its own transaction context.
To fully understand a user's NFT ownership, we leverage the transaction histories of all the user's NFTs. 
We construct a \textit{NFT-User graph} to model ownership changes in NFTs and train an extractor for the graph to extract NFT transaction context of each user.
From the NFT-User graph, we first obtain the transaction context of each NFT by aggregating its transaction history. Finally, to get a representation of the user's NFT transaction context, we aggregate the transaction contexts of all of their owned NFTs. 

\subsubsection{NFT-User graph construction}

We construct an undirected graph $G(U,N,E_T)$.
There are two types of nodes: user nodes $U$, NFT nodes $N$, and the $u$ nodes and $n$ nodes can be connected with an attributed edge $e \in E_T$, called \textit{NFT ownership attributes}.
% where $S_T$ is the size of the feature space for each edge. 
Additionally, $N_e(n)$ is the node $n$’s one-hop neighbors (users who traded NFT $n$), $N_e(n) = \{u' \in U | (u',n) \in E_T \}$, and $N_e(u)$ is the set of NFT nodes in node $u$’s one-hop neighbors, $N_e(u) = \{n' \in N | (n',u) \in E_T \}$. 
% Each edge has seven NFT ownership attributes (Section \ref{s:nft_tx_att}).
% Note that we remove NFT nodes connected to less than three user nodes since it is difficult to capture regular transaction patterns.

\subsubsection{NFT transaction context extraction}
% To fully understand a user's NFT ownerships, we leverage the transaction histories of all the user's NFTs. 

A transaction context $h^N$ of an NFT node $n$ is updated based on attributed edges connected to, denoted as $T_n$ = $\{t_{u_1}, t_{u_2}, \dots, t_{u_m} \}$, where $u_i$ ($1\le i \le m$) is one of $N_e(n)$, and $m$ is the size of $N_e(n)$.
Then, we aggregate them with convolution layer as follows :
$$h_n^N = \sigma  \left( W^N \cdot AGG_N(T_n) \right)$$
where $\sigma$ is an activation function, and $W^N$ is the trainable matrix for learning. We use mean-pooling as an aggregation function to represent the transaction context of each NFT.

Next, a user representation $h^U$ is updated by all neighboring NFT nodes' transaction context vectors.
For a user $u$, the representation $h_{un}^U$ of each NFT $n \in N_e(u)$ is obtained as follows. 

$$h_{un}^U = W^U \cdot concat(t_{un}, h_{n}^N)$$

where $t_{un}$ is  $u$'s ownership of $n$, and $h_{n}^N$ is the transaction context vector of $n$. We concatenate the two and apply a linear transformation with the learnable matrix $W^U$.

Finally, we integrate all of the NFT transaction context vectors $\{h_{n_1}^N, h_{n_2}^N, \dots, h_{n_v}^N \}$ ($v$ is the size of $N_e(u)$). To be sensitive to irregular NFT transaction patterns, we implement attention mechanisms to combine each NFT ownership vector. Specifically, we adapt a multi-head graph attention operation to reduce the effects of noise.

$$z_{un} = \mathrm{LeakyReLU}(a \cdot h_{un}^U)$$
$$h_{u}^U = \sum_{n \in N_e(u)} \alpha_{un} z_{un}$$

where $z_{un}$ is an attention score calculated by taking the dot product of learnable weight $a$ and applying $\mathrm{LeakyReLU}$. $\alpha_{un}$ is the normalized attention score using the softmax function. 
% $h_{uk}^U$ is calculated by the \textit{k-th} attention head, where $K$ is the number of heads. 
Lastly, we compute the final representation by averaging the attention head outputs.

\noindent\textbf{Training.}
The final representation feeds into the classification layer for classification. 
The extractor updates the trainable parameters to better learn features that distinguish drainers from regular users.
We use cross-entropy loss as the loss function.

\subsection{Social Context Extractor}
\label{s:social_context}
Drainers have distinct motivations for engaging in NFT trading compared to regular users, leading to the formation of unique social connections.
Most drainers choose to liquidate through affiliated accounts. Moreover, some affiliated accounts are used by multiple drainers, which suggests a close relationship between the co-users. Thus, the relationships between users are essential to detect drainers. To model user interactions, we construct a \textit{User graph} and use it to train an extractor to learn the social context of users.

\subsubsection{User graph construction}
\label{s:uer_graph_construction}
Trading between users can be constructed as a graph $G(U,E,R,X_U)$, where $U$ is the set of user nodes, and $E$ is labeled edges $(u_i, r, u_j)$, where $r \in R = \{sale, gift\}$ is a relation type.  
Each user node has 19-dimensional user node attributes, $X_U$.
If $u_i$ transfers an NFT to $u_j$, then $u_i$ and $u_j$ are connected with an edge $e \in E$ with a label \textit{gift}. 
% In the user graph, a node has 
% $X_U \in \mathbb{R}^{{\mid U \mid}\times{\mid S_U \mid}}$, where $S_U$ is the feature space size.

\subsubsection{Social context extraction}
Understanding user relationships in the NFT ecosystem relies heavily on transaction types. Therefore, we utilize transaction types as relational information between users. This approach allows us to capture relational dependencies and acquire more meaningful representations.

To accomplish this, we employ the R-GCN model (Relational Graph Convolutional Network)~\cite{schlichtkrull2018modeling}, which is specifically designed to handle graph structures with relational data. The R-GCN model has demonstrated strong performance across various tasks and is well-suited for our purposes.
% Transaction types are important to understand user relationships in the NFT ecosystem. To consider transaction types between users, we use the R-GCN~\cite{schlichtkrull2018modeling} model, where edges can represent different relations.

From the user graph, we obtain a representation vector of a user node $u$ updated by its neighboring user nodes. The propagation at $(l+1)$-th layer of R-GCN with $L$ layers is as follows:

$$h_u^{l+1}=\sigma \left(W^{l} h_u^{l} \cdot \sum_{r \in R }  \mathrm{AGG_U} (\frac{1}{c_{u,r}} W_r^{l}h_{v}^{l}), {\forall v \in N(u)_r} \right)$$

where $\sigma$ is an activation function, and $W^{l}$ is a learnable matrix shared among all nodes at $l$-th sub-layer. 
$N(u)_r$ is neighboring user nodes under relation $r$ of $u$. 
We use mean-pooling as the aggregation function $\mathrm{AGG_U}$.

\noindent\textbf{Training.} We train this module in the same process as the NFT transaction context extractor by feeding the outputs into the classification layer and using cross-entropy loss.

\subsection{Drainer Classifier}
\label{s:drainer_classifier}
Following the above operations, we obtain three types of features representing users: (1) NFT transaction context representation, (2) social context representation, and (3) user node attributes. 
We concatenate the three representations together to create our final representation, integrating comprehensive information learned from our graphs.
In order to learn the differences between drainers and regular users, we feed the final representation to a classifier layer.

% $$y = classifier(concat(h_u^I, h_u^R, h_u^S))$$
We choose a support vector machine (SVM)~\cite{hearst1998support} as our classifier layer. 
SVM is a supervised machine learning model that uses classification algorithms.
SVM has the advantage of reducing the chances of model overfitting, making the model highly stable. SVM is also powerful to deal with high dimensional features.
% We choose LightGBM~\cite{ke2017lightgbm} as our classifier layer. LightGBM is based on Gradient Boosted Machines (machine learning algorithms that use an ensemble of weak learners to improve performance).
% LightGBM uses decision trees for various tasks, such as classification and regression tasks.

\begin{table}[t]
    \centering
    \caption{Dataset statistics for training and evaluation. Ratio refers to the ratio of drainers to sampled regular users. For evaluation datasets, each number is the average value over 5 runs. }
    \label{t:sum_datasets}
    \footnotesize
    \begin{tabular}{lrrrrr}
        \toprule
        \multicolumn{2}{l}{Dataset} & Ratio & \makecell[c]{\# central \\ nodes }& \makecell[c]{\# total\\ nodes} & \# transactions\\
        \midrule
        Training & $D_0$ & 1:80 & 52,245 & 2,010,384.0 & 24,745,525.0 \\
        \midrule
        \multirow{4}{*}{Evaluation} 
        & $D_1$ & 1:10  & 6,006 & 2,087,436.0 & 28,375,070.6 \\
        & $D_2$ & 1:100  & 55,146 & 2,743,003.4 & 41,384,504.8 \\
        & $D_3$ & 1:1000  & 546,546 & 3,179,105.4 & 45,289,602.6 \\
        \bottomrule
    \end{tabular}
\end{table}
\section{Experiments}
\label{sec:experiment}

To validate our model in different aspects, we present several empirical evaluations in this section. Specifically, we seek to answer the following research questions: 

- How effective is \model{} in detecting NFT drainers? 

- How does each component affect performance?

- How robust is our model against evasion attacks?

\begin{table*}[t]
\caption{The results of experiments averaged over 5 runs on datasets $D_1 \sim D_3$. Pre., Rec., F1., and FP/TP mean precision, recall, F1 score, and the number of false positive/true positive, respectively. }
% \textit{ Node attr.} denotes the user attributes used in the User graph of \model{}.}
\label{t:performance}
\centering
\ra{1.03}
\footnotesize
\begin{tabular}{cr c@{\hspace{9pt}}c@{\hspace{9pt}}c@{\hspace{9pt}}r| 
                c@{\hspace{9pt}}c@{\hspace{9pt}}c@{\hspace{9pt}}r| 
                c@{\hspace{9pt}}c@{\hspace{9pt}}c@{\hspace{9pt}}r}
    \toprule
    \multirow{3}{*}{Model} 
    & Dataset (ratio)  
    & \multicolumn{4}{c}{$D_1$ (1:10)} 
    & \multicolumn{4}{c}{$D_2$ (1:100)} 
    & \multicolumn{4}{c}{$D_3$ (1:1000)}\\
    \cmidrule{2-14}
    & Metrics   
    & Pre. & Rec. & F1 & \multicolumn{1}{c}{FP/TP}
    & Pre. & Rec. & F1 & \multicolumn{1}{c}{FP/TP}
    & Pre. & Rec. & F1 & \multicolumn{1}{c}{FP/TP}\\
    \midrule
    \multirow{3}{*}{\makecell[c]{Feature\\based}}
    & Ether features
    & 0.875 & 0.227 & 0.361 &  12.2/85.0
    & 0.429 & 0.227 & 0.297 &  113.2/85.0
    & 0.072 & 0.227 & 0.109 & 1096.6/85.0 \\
    & E-GCN features 
    & 0.838 & 0.104 & 0.185 & 10.0/51.0
    & 0.334 & 0.104 & 0.159 & 102.4/51.0
    & 0.047 & 0.104 & 0.064 & 1045.4/51.0 \\
    & \model{} user features  &
     0.976 & \underline{0.618} & \underline{0.757} &7.4/302.4 & 
     0.779 & \underline{0.618} & \underline{0.689} & 86.2/304.2 & 
     0.277 & \underline{0.627} & 0.385 & 886.4/340.2  \\
    \midrule
    \multirow{7}{*}{\makecell[c]{Graph\\based}}
    & Node2Vec & 
    0.000 & 0.000 & 0.000 & 0.0/0.0 &
    0.000 & 0.000 & 0.000 & 0.0/0.0 &
    0.000 & 0.000 & 0.000 & 0.0/0.0 \\
    & E-GCN & 
    0.832 & 0.037 & 0.071 & 0.0/1.0 & 
     0.349 & 0.037 & 0.067 & 2.0/1.0 & 
     0.055 & 0.037 & 0.044 &28.6/1.0 \\
    & E-GAT&
    0.933 & 0.010 & 0.020 & 0.4/5.0 &  
     0.825 & 0.010 & 0.020 & 1.2/5.0 &  
     0.256 & 0.009 & 0.018 & 15.0/5.0 \\
    & E-GraphSAGE &
    0.980 & 0.157 & 0.271 & 1.6/77.0 &  
    0.867 & 0.157 & 0.265 & 12.0/77.4 &  
    0.435 & 0.157 & 0.231 & 111.2/85.4 \\
     
     & N-GCN & 
      0.838 & 0.103 & 0.183 & 9.8/50.2 & 
     0.351 & 0.103 & 0.159 &  93.8/50.6 & 
     0.057 & 0.102 & 0.073 &923.0/55.4 \\

    & N-GAT &
     0.982 & 0.411 & 0.580 & 3.8/201.4 & 
     0.811 & 0.411 & 0.546 & 47.4/202.6 & 
    0.323 & 0.415 & 0.363 & 471.2/225.2 \\

    & N-GraphSAGE &
     \underline{0.987} & 0.569 & 0.722 &  3.6/278.4 & 
     \underline{0.860} & 0.569 & 0.685 & 45.8/280.2 & 
     0.416 & 0.579 & \underline{0.484} & 441.2/314.4 \\ 
    \midrule
    \multicolumn{2}{c}{\model{}} & 
     \textbf{0.989} & \textbf{0.622} & \textbf{0.763} & 3.4/304.4 & 
     \textbf{0.878} & \textbf{0.621} & \textbf{0.727} & 42.8/306.0 & 
     \textbf{0.448} & \textbf{0.631} & \textbf{0.523} & 422.8/342.4 \\ 
    \bottomrule
\end{tabular}
\end{table*}

\subsection{Datasets}
\label{sec/dataset}
\diff{In our experiments, we utilize} a dataset consisting of NFT transaction data and accounts identified as NFT drainers.

\noindent\textbf{Potential False Negative Filtering:} 
In Section~\ref{sec:measurement}, we observe that drainers often gift stolen NFTs to \textit{affiliated users}, and that some affiliated users receive NFT gifts from multiple drainers.
From this observation, it can be suggested that the other accounts that gift NFTs to known affiliated users are also highly suspected to be related to drainers. 
However, these accounts, while suspicious, cannot be determined with certainty as drainers themselves.
Therefore, we choose to exclude these suspicious accounts as well as affiliated users from the regular user category. 
% We filtered out both the affiliated users themselves and the accounts that gifted NFTs to affiliated users.

\noindent\textbf{Training Dataset Construction:}
To create our training dataset, we first \diff{gathered accounts that engaged in transactions from} January 1, 2022, and July 31, 2022. 
\diff{This was comprised of 3,137,221 accounts, with 645 of them being identified as drainer accounts.}
Due to the highly imbalanced ratio of regular users to drainers, directly using this data for training would not be effective.
Therefore, \diff{we used two sampling strategies to select regular users for the training set.}

First, we excluded accounts that have low activity.
Specifically, we removed two categories of accounts: 1) accounts that had never received NFTs from \diff{other users} and 2) accounts with zero active time.
\diff{From the remaining 1,355,811 accounts after removal, we sampled 45,150 regular users (at a 1:70 ratio) to include in our training dataset.}

Secondly, we additionally sampled ``heavy" regular users\diff{, specifically those with over 50 transactions.}
This strategy aims to address the potential model bias towards transaction quantity\diff{, since regular users tend to partake in fewer transactions than drainers} (see Appendix~\ref{apx:tx_dist}). Thus, we select an additional 6,450 heavy regular users (at a 1:10 ratio) into the training dataset.

Our final training dataset comprises 645 drainers and 51,600 (45,150+6,450) regular users. 
The specific ratios used in the sampling processes were empirically selected through experimental evaluation.
\diff{The ideal number of regular users can depend on the class ratio in the evaluation dataset. For a deeper exploration of how the class ratio in the training dataset impacts results, refer to Section~\ref{sec:discussion}.}

\noindent\textbf{Evaluation Dataset Construction:}
For the evaluation dataset, we \diff{selected} accounts that \diff{engaged in} at least one transaction between August 1, 2022, and December 31, 2022. \diff{This yielded a total of 1,723,465 accounts, of which 546 were identified as drainers.} 
Note that \diff{our dataset utilizes transaction records of these accounts from January 1, 2022 to December 31, 2022. Given the notable imbalance between drainers and regular users in the dataset, it is critical to evaluate our model under various scenarios. To achieve this, we created three separate datasets by adjusting the proportion of regular users in our tests to be 10, 100, and 1000 times the number of drainers. 
By doing so, we aim to understand our model's performance under varying levels of class imbalance. 
While higher ratios of regular users offer a representation more in line with real-world distributions, it is more lenient towards false negatives, which translates to drainer accounts that remained undetected.}

\diff{To construct each dataset, we use the selected accounts as central nodes and include first and second-order neighbor nodes. Note these neighbor nodes are used only to enrich the graph, and are not used for training or evaluation.
The dataset statistics are summarized in Table~\ref{t:sum_datasets}.}

\diff{\subsection{Baselines}}

% \noindent\textbf{Baseline Models:} 

As baselines, we use methods that effectively detect Ethereum phishing accounts.
In addition, we also use other widely used graph-based models because \model{} is a graph-based model.
The baselines can be divided into two categories: Feature-based and graph-based.

\vspace{5pt}

\textbf{Feature-based}:
\begin{itemize}
    \item \textbf{Ether features}~\cite{chen2020phishing} use 119-dimensional statistical features previously used for Ethereum phishing account detection. The features mainly consist of first-order neighbor information. 
    \item \textbf{E-GCN features}~\cite{EGCN} are the initial node features used in E-GCN, a method proposed to detect Ethereum phishing accounts.
    \item \textbf{\model{} user features} are the initial node features used in our User-graph.
\end{itemize}

\textbf{Graph-based}:

\begin{itemize}
    \item \textbf{Node2Vec}~\cite{grover2016node2vec} is a graph node embedding method based on random walks. 
    \item \textbf{E-GCN}~\cite{EGCN} applies a Graph Convolutional Network (GCN) to detect Ethereum phishing accounts. 
    \item \textbf{GAT}~\cite{GAT} is a widely used GNN model. It learns node representations by aggregating neighbor nodes with an attention mechanism.
    \item \textbf{GraphSAGE}~\cite{graphsage} is another GNN variant. It learns node representations by aggregating sampled neighbor node features.
\end{itemize}

To evaluate the effectiveness of our features, we implement each graph-based baseline twice: 
once using E-GCN features (denoted by the "E-" prefix) and once using \model{} user features (denoted by "N-" prefix).

\begin{table*}[t]
\caption{
The results of ablation experiments averaged over 5 runs.
The most impactful features of each attribute are highlighted in color. 
(*) indicates a set of features involving in and out directions, and (**) indicates a set of features involving all transaction types.}
\label{t:ab_study2}
\centering
\ra{1.03}
\footnotesize
\begin{tabular}{cr ccc| ccc| ccc }
    \toprule
    \multirow{2}{*}{Type} 
    & Dataset (ratio)  
    & \multicolumn{3}{c}{$D_1$ (1:10)} 
    & \multicolumn{3}{c}{$D_2$ (1:100)} 
    & \multicolumn{3}{c}{$D_3$ (1:1000)} \\
    \cmidrule{2-11}
    & Removed \textit{feature} group 
             & Pre. & Rec. & F1 
             & Pre. & Rec. & F1  
             & Pre. & Rec. & F1  \\
    \midrule 
   & Holding time &
    0.989 & 0.605 & 0.751 &
    0.878 & 0.605 & 0.717 &
    0.440 & 0.609 & 0.511 \\
   & Transaction type* &
    0.986 & 0.634 & 0.772 &
    0.867 & 0.634 & 0.732 &
    0.426 & 0.640 & 0.511 \\
    \rowcolor{blue!15}
    \cellcolor[rgb]{ 1,  1,  1}
   & Price* &
    0.984 & 0.618 & 0.759 &
    0.867 & 0.617 & 0.721 &
    0.425 & 0.627 & 0.506 \\
    \rowcolor{blue!40}
    \cellcolor[rgb]{ 1,  1,  1}
    & Avg. price \& holding time  &
    0.988 & 0.595 & 0.743 &
    0.869 & 0.595 & 0.706 &
    0.429 & 0.600 & 0.500 \\
    &  In-transaction type \& In-price &
    0.986 & 0.616 & 0.758 &
    0.865 & 0.616 & 0.719 &
    0.428 & 0.624 & 0.507 \\
    \rowcolor{blue!30}
    \multirow{-6}{*}{NFT ownership edge attributes}
    \cellcolor[rgb]{ 1,  1,  1}
    & Out-transaction type \& Out-price  &
    0.986 & 0.618 & 0.759 &
    0.865 & 0.617 & 0.720 &
    0.417 & 0.627 & 0.501 \\

    \midrule
    \rowcolor{blue!15}    
    \cellcolor[rgb]{ 1,  1,  1}
    \multirow{7}{*}{User node attributes}   &
    Active timespan &
    0.986 & 0.609 & 0.753 &
    0.850 & 0.609 & 0.710 &
    0.392 & 0.618 & 0.480 \\
    & Gift-in ratio &
    0.988 & 0.620 & 0.762 &
    0.864 & 0.620 & 0.722 &
    0.414 & 0.627 & 0.499 \\
    & Out-in ratio &
    0.987 & 0.624 & 0.764 &
    0.873 & 0.624 & 0.727 &
    0.427 & 0.634 & 0.510 \\
    & \# transactions** &
    0.990 & 0.558 & 0.714 &
    0.887 & 0.558 & 0.685 &
    0.467 & 0.569 & 0.513 \\
    \rowcolor{blue!40}
    \cellcolor[rgb]{ 1,  1,  1}& 
    \# collections** &
    0.985 & 0.595 & 0.742 &
    0.842 & 0.595 & 0.697 &
    0.376 & 0.604 & 0.464 \\
    \rowcolor{blue!30}
    \cellcolor[rgb]{ 1,  1,  1}& 
    \# neighbors** &
    0.985 & 0.597 & 0.744 &
    0.842 & 0.597 & 0.699 &
    0.375 & 0.608 & 0.464 \\
    & Freq. of gift-in \& sell &
    0.987 & 0.612 & 0.755 &
    0.871 & 0.611 & 0.718 &
    0.431 & 0.622 & 0.509 \\
    \midrule

    \multicolumn{2}{c}{\model{}} &
     \textbf{0.989} & \textbf{0.622} & \textbf{0.763} & 
     \textbf{0.878} & \textbf{0.621} & \textbf{0.727} & 
     \textbf{0.448} & \textbf{0.631} & \textbf{0.523}  \\ 
    \bottomrule
\end{tabular}
\end{table*}
\subsection{Experimental Results}
\diff{For implementation details, refer to Appendix~\ref{apx:implementation}. Table~\ref{t:performance} shows that our model outperforms the baselines in all evaluation metrics, which proves the effectiveness of \model{} for NFT drainer detection.}
% Despite the substantial class imbalance, our model achieves an F1-score exceeding 0.5, highlighting its efficacy in addressing the detection task.

\diff{\model{} outperforms standard feature-based methods since it benefits from the NFT-specific features and high-level representations obtained from extractors.}
Many NFT-specific features can be instrumental in distinguishing NFT drainers from regular users but are ignored in the existing methods. 
\diff{On the other hand, previous approaches use features that may not fully apply to NFT trading.}
\diff{For instance, a significant feature} in Ethereum phishing account detection~\cite{chen2020phishing} \diff{is whether an account has mixing services as neighbors, since phishers tend to liquidate with such services.}
\diff{However, since NFTs cannot go through mixing services, this feature is inapplicable for analyzing NFT trading.}

One surprising finding is that graph-based methods without our NFT-specific attributes \diff{(E-GCN, E-GAT, and E-GraphSAGE)} perform poorly compared to the feature-based methods. 
For instance, \diff{solely relying on the E-GCN features yields better results than the complete E-GCN method itself.}
\diff{Furthermore,} both E-GCN and E-GAT \diff{tend to classify the majority of users as non-malicious, leading to an alarmingly} low recall.
However, \diff{this trend changes when we embed} NFT-specific attributes \diff{into these} graph-based methods.
\diff{By doing so, the performance of the modified graph-based methods} (N-GCN, N-GAT, and N-GraphSAGE) improved significantly, highlighting the importance of our feature engineering. 
In addition to the NFT-specific attributes, \model{} takes into account transaction types and utilizes NFT transaction context extracted from the NFT-User graph. This results in a multi-faceted representation that allows it to outperform other graph-based methods.

\diffnote{MR2} \diff{\model{} outperforms the baselines in NFT drainer detection, but we observe that precision and F1-score decrease as the dataset size increases.
While we have addressed potential false negatives when constructing datasets, unreported drainers may still exist among what we categorize as regular users.  
Naturally, the decrease in precision might have come from correctly classifying unreported drainers as drainers. 
We will discuss the unreported drainers in Section~\ref{sec:disc_fp}.

}
% \vspace{0.1}
% \input{figure/ablation_study.tex}
\begin{table}[t]
\caption{
The results of ablation experiments \diff{on $D_1$ and $D_3$ datasets} averaged over 5 runs.}
\label{t:ab_study1}
\centering
\ra{1.03}
\footnotesize
% \resizebox{\columnwidth}{!}{%
\begin{tabular}{l c@{\hspace{6pt}}c@{\hspace{6pt}}c@{\hspace{6pt}}| c@{\hspace{6pt}}c@{\hspace{6pt}}c@{\hspace{6pt}}}
    \toprule 
    Dataset (ratio)  
    & \multicolumn{3}{c}{$D_1$ (1:10)} 
    & \multicolumn{3}{c}{$D_3$ (1:1000)} \\
    \midrule
    Removed & Pre. & Rec. & F1 
     & Pre. & Rec. & F1  \\
    \midrule 
    \model{} user features &
    0.985 & 0.599 & 0.745 &
    0.410 & 0.607 & 0.489 \\
    NFT transaction context &
    0.989 & 0.585 & 0.735 &
    0.450 & 0.588 & 0.510 \\
    Social context&
    0.984 & 0.591 & 0.739 &
    0.391 & 0.595 & 0.472 \\
    Relation in SCE &
    0.980 & 0.589 & 0.736 &
    0.327 & 0.594 & 0.422 \\
    \midrule
    \model{} &
    0.989 & 0.622 & 0.763 &
    0.448 & 0.631 & 0.523 \\
    \bottomrule
\end{tabular}
% }

\end{table}
\vspace{-0.2cm}

\subsection{Ablation Study}
\label{sec:abstudy}
In this section, we \diff{delve into understanding how individual components within \model{} impact} its overall performance.
\diff{We first assess the influence of each representation by removing them one-by-one.}
\diff{Specifically, we assess the representations that stem from} our feature engineering, \firstmodule{} (TCE), and \secondmodule{} (SCE).
In addition, to \diff{understand} the importance of considering transaction types in the User graph, we \diff{experiment with a} version of \model{} \diff{that employs} GCN instead of R-GCN.

\diff{The experimental results, presented in Table~\ref{t:ab_study1}, (and results on $D_2$ in Appendix~\ref{apx:ab_study}) show that optimal performance is achieved when all components are integrated.}
\diff{Exclusion of the NFT transaction context leads to a marked drop in recall. This underscores its importance in identifying drainers, which other components might miss.}
Additionally, the removal of the social context representation yields the lowest precision \diff{of all components}.
\diff{This suggests its pivotal role} in preventing \model{} from misclassifying regular users as drainers. 
\diff{Excluding relations in the component leads to a drop in} performance, \diff{especially} as the number of regular users increases. 
This highlights the importance of transaction types for \diff{identifying} NFT drainers.

\begin{table*}[t]
\diffnote{MR1}
\ifdiff
\begin{mdframed}[backgroundcolor=blue!10]
\fi
\caption{
The results of evasion attacks on $D_1$ and $D_3$ datasets averaged over 5 runs. 
We re-trained the classifier layer with 3\% of evasion attackers and evaluated remaining evasion attackers on datasets $D_1'$ and $D_3'$. The result for other values of $X$ refers to Appendix~\ref{apx:evasion_attack_X}.}
\label{t:evasion_new}
\centering
\ra{1.03}
\footnotesize
\begin{tabular}{ccc ccc| ccc | ccc | ccc}
    \toprule
    & Dataset (ratio) &
    & \multicolumn{3}{c}{$D_1$ (1:10)} 
    & \multicolumn{3}{c}{$D_1'$ (1:10)}
    & \multicolumn{3}{c}{$D_3$ (1:1000)}
    & \multicolumn{3}{c}{$D_3'$ (1:1000)}\\
    \cmidrule{2-15}
    & L & X
    & Pre. & Rec. & F1 
    & Pre. & Rec. & F1 
    & Pre. & Rec. & F1 
    & Pre. & Rec. & F1 \\
    \midrule
    \multirow{3}{*}{Attack1}
    & 10 &&
    0.970 & 0.481 & 0.643 &
    0.981 & 0.542 & 0.698 &
    0.278 & 0.484 & 0.353 &
    0.263 & 0.547 & 0.356 \\
    & 30 & N/A&
    0.962 & 0.386 & 0.551 &
    0.979 & 0.507 & 0.668 &
    0.238 & 0.393 & 0.296 &
    0.254 & 0.517 & 0.341 \\
    & 50 && 
    0.959 & 0.354 & 0.517 &
    0.979 & 0.497 & 0.660 &
    0.221 & 0.358 & 0.273 &
    0.251 & 0.507 & 0.336 \\
    \midrule
   \multirow{3}{*}{Attack2}
    & 10 &&
    0.966 & 0.349 & 0.513 &
    0.979 & 0.601 & 0.744 &
    0.218 & 0.356 & 0.270 &
    0.269 & 0.607 & 0.373 \\
    & 30 & N/A&
    0.940 & 0.192 & 0.319 &
    0.980 & 0.635 & 0.771 &
    0.132 & 0.194 & 0.157 &
    0.277 & 0.639 & 0.387 \\
    & 50 && 
    0.919 & 0.139 & 0.241 &
    0.981 & 0.663 & 0.791 &
    0.097 & 0.137 & 0.114 &
    0.287 & 0.669 & 0.401  \\
    \midrule
    \multirow{3}{*}{Attack3}
    & 10 &  &
    0.866 & 0.110 & 0.195 &
    0.966 & 0.574 & 0.719 &
    0.081 & 0.115 & 0.095 &
    0.220 & 0.591 & 0.320  \\
    & 30 & 60 &
    0.852 & 0.098 & 0.176 &
    0.965 & 0.625 & 0.758 &
    0.074 & 0.104 & 0.086 &
    0.222 & 0.635 & 0.328 \\
    & 50 &  &
    0.873 & 0.114 & 0.202 &
    0.970 & 0.644 & 0.774 &
    0.082 & 0.118 & 0.097 &
    0.264 & 0.648 & 0.374 \\
    \midrule
    \multirow{3}{*}{Attack4}
    & 10 &  &
    0.551 & 0.020 & 0.039 &
    0.952 & 0.425 & 0.587 &
    0.017 & 0.023 & 0.019 &
    0.171 & 0.474 & 0.251 \\
    & 30 & 60 &
    0.426 & 0.012 & 0.024 &
    0.956 & 0.563 & 0.709 &
    0.010 & 0.013 & 0.012 &
    0.183 & 0.587 & 0.278\\
    & 50 &  &
    0.430 & 0.012 & 0.024 &
    0.961 & 0.634 & 0.764 &
    0.011 & 0.014 & 0.012 &
    0.207 & 0.651 & 0.314 \\
    \midrule
    \multicolumn{3}{c}{\model{}} & 
    \textbf{0.989} & \textbf{0.622} & \textbf{0.763} & 
    \textbf{0.989} & \textbf{0.622} & \textbf{0.763} & 
    \textbf{0.448} & \textbf{0.631} & \textbf{0.523} &
    \textbf{0.448} & \textbf{0.631} & \textbf{0.523} \\

    \bottomrule
\end{tabular}
\ifdiff
\end{mdframed}
\fi

\vspace{-0.3cm}
\end{table*}

We also examine the \diff{influence} of \textit{NFT transaction edge attributes} and \textit{user node attributes} \diff{on \model{}'s performance}. We grouped features that represent similar concepts together. For example, the number of transactions for each transaction type (5) was grouped into one, \textit{the number of transactions}. 
Then, we trained and evaluated \model{} on the graph constructed without each group.

In Table~\ref{t:ab_study2}, \diff{critical components in the TCE component include} \textit{average information}, \textit{out-transaction type \& out-price},  and \textit{price}.
Notably, \diff{discarding} average information for each NFT \diff{significantly impairs performance.}
\diff{Its inclusion aids the model in differentiating between regular and non-regular transactions based on NFT transaction patterns.} 
The price and out-transaction information also play a crucial role in detecting drainers, capturing their distinctive pattern of selling NFTs at a lower price.
\diff{Another significant observation is the increased F1-score in datasets $D_1$ and $D_2$ upon removal of \textit{transaction type}}. 
However, this improvement was offset by a \diff{drop} in precision, resulting in poor performance in F1-score on dataset $D_3$.
This highlights the importance of considering transaction type to avoid misclassifying regular users as drainers, particularly as the number of users increases.

The user node attributes exhibited a greater influence on performance compared to the NFT transaction edge attributes.
\diff{The most critical attributes are} \textit{the number of collections}, \textit{the number of neighbors}, and \textit{active timespan}. 
This aligns with \diff{drainer behavior of trying} to steal NFTs from a wide range of users. Consequently, drainers tend to engage in trades involving numerous NFT collections and interact with a large number of neighbors within \diff{a brief period.}
% Removing \textit{out-in ratio} results in a higher recall but lower precision compared to \model{}. 
% This suggests that drainers have unique characteristics not present in regular users, and they help \model{} avoid misclassifying regular users as drainers.

% \diff{In summary, our study underlines the significance of each representation and validates the efficacy of our feature engineering in enhancing \model{}'s detection capabilities.}

% \input{table/evasion1.tex}

\subsection{Robustness}

\label{sec:robustness}
If drainers notice that they are monitored by \model{}, they may purposely change their trading patterns to avoid detection. 
\diff{Therefore, we discuss and evaluate to what extent an adversary can deceive \model{}.}

\diffnote{MR1}\diff{
\subsubsection{Evasion attacks}
The following are the environmental assumptions and constraints for evasion attacks.

\textbf{Assumptions.} 
\model{} closely monitors every transaction and swiftly alerts marketplaces when a new drainer is detected.
Meanwhile, victims can directly report the drainer if they realize they've been compromised.
On confirming a drainer's malicious activities, marketplaces instantly freeze their NFT trades to prevent further sales of stolen assets.
For a comprehensive version of the actual usage scenario of \model{}, refer to Section~\ref{sec:discussion}.

\textbf{Constraints.} 
To benefit from stolen NFTS, drainers must rapidly sell the NFTs at lower prices, particularly before marketplace bans.
Also, the selling processes for stolen NFTs are influenced by both regular users and drainers, not solely by the drainers.
Thus, the latter cannot afford to alter specific parameters related to selling of stolen assets - like \textit{holding time, out-transaction type, out-price,} and \textit{frequency of selling}.
Under the assumptions and constraints, drainers essentially have two avenues for evasion.

\textbf{1. Utilizing Multiple Accounts.}
A drainer could deploy multiple accounts. 
In fact, using affiliated users for selling NFTs can be a part of this type of attack.
While \model{} can spot drainers making sales through affiliated users, their methods can become more sophisticated.
For example, during liquidation, a drainer might use multiple auxiliary accounts to trade an NFT cyclically, maintaining its original value, before selling it to a regular user at a reduced price.
However, it is commonplace for users to operate multiple wallet accounts for better asset management. 
For instance, an individual could relocate assets from other accounts to a different account for sale.
This can pose a challenge for \model{} as it might struggle to differentiate between this layered attack strategy and legitimate trades. 
Nonetheless, these attackers are still pressured to liquidate their NFTs across multiple sales in a restricted timeframe to evade marketplace bans, which results in a conspicuous spike in trading volume.
While \model{} currently does not provide countermeasures for this layered attack,
a security operator (or wash trading detection system) might grow suspicious of an unexpected trading volume surge of particular NFTs, even if the individual transactions do not seem malicious.
 
\textbf{2. Using a Single Account.}
A drainer can also mask its activity by changing the trading pattern of the drainer account itself without the use of additional accounts. 
It can make a series of low-value noise transactions to alter its user node attributes, NFT owner edge attributes, and inter-relationships in the graphs. 
Thus, we introduce four types of attacks designed to change these attributes and graphs significantly within the given constraints.
}

\textit{\textbf{Attack 1. Mint NFTs.} }
The easiest way that drainers can engage in innocuous-appearing NFT transactions is by minting NFTs. 
\diff{With this attack, drainers can alter numerous user node attributes, including gift-in and out-in ratio.}

\diff{
\textit{\textbf{Attack 2. Increase Active timespan.} }
An active timespan is one of the important features distinguishing drainers from regular users.
Drainers can easily increase their active timespan by engaging in a transaction before initiating draining activity.
}

\textit{\textbf{\diff{Attack 3}. Send Ether to victim.} }
Drainers can hide their activities by sending Ether to victims after stealing NFTs. 
This way, \diff{draining} is recorded as a sale instead of a gift. 
\diff{This attack not only changes most of the user node and NFT ownership edge attributes but also alters both associated graphs.}

\diff{
\textit{\textbf{Attack 4. Combination of three attacks.}} This is the combination of all three attacks.
}

\subsubsection{Evaluation}
\diff{To evaluate \model{}'s detection capabilities against evasion tactics, we adjusted previous evaluation datasets.
These modifications were guided by the specific attack strategy and its attack level ($L$), where $L \in \{10, 30, 50\}$.}

For Attack 1, we increased the number of minted NFTs by $L$\% of gifted-in NFTs. 
\diff{For Attack 2, the active timespan was extended by $L$\%.}  
\diff{For Attack 3,} we changed $L$\% of gifting-in transactions to buying transactions by sending X\% of the average sale price of each NFT to those victims, where $X \in \{1, 10, 60\}$. 
\diff{Lastly, for Attack 4, we integrated the tactics of Attack 1 and Attack 2, both at level 50, with Attack 3.}
We set $X$ to be less than 60 because drainers typically sell stolen NFTs at 40\% below their average sale price \diff{(based on our findings in Section~\ref{sec:measurement})}.

\diff{The experimental results, presented in Table~\ref{t:evasion_new}, (with results on $D_2$ in Appendix~\ref{apx:evasion_attack}) show that as the attack strategies progress from Attack 1 to Attack 4, the system's performance is increasingly compromised.
Attack 2 poses a greater threat than Attack 1, suggesting that a short active timespan was a critical trait of drainers.
Nevertheless, Attack 1 and Attack 2 lag behind even the lowest intensity of Attack 3 and Attack 4. 
This stems from their inability to modify the user and NFT relationships in the graphs.
}

On the other hand, \diff{Attack 3 and Attack 4} poses a significant challenge to our system's detection capabilities.
This is because attackers adopting \diff{Attack 3} deviate from the definition of drainers we used in this study (accounts that steal NFTs).
Since the model was trained to identify accounts that steal NFTs, it is not optimized for detecting cases in which attackers buy NFTs at lower prices.
As a result, these attackers are considered an unseen data type within our model, causing difficulties in detecting them. 
\diff{Table~\ref{apx:evasion_attack_X} (in Appendix~\ref{apx:evasion_attack_X}) shows that Attack 3} is more effective in evading the detector when more Ether is sent to victims.
However, this comes at a cost for the attacker since each draining operation will incur additional costs.
Despite the effectiveness of \diff{these attacks}, it is critical to highlight the resilience of our system; this method fails to fully deceive the updated system with the defense method (which will be introduced below).

\subsubsection{Defense}
\label{sec:defense}
\diff{
A proactive defense against evasive attacks entails periodically refining \model{} to recognize emerging drainer patterns.
}
We update \model{} in a simplified manner, re-training only the last layer (SVM), while leaving two extractors intact. 
For the evaluation, we trained the SVM using a newly augmented dataset \diff{that combines the prior} training dataset along with an additional 3\% of evasion attackers. 
\diff{This updated model was then tested against the remaining evasion attackers.} 
We trained and evaluated using a consistent ratio of drainers to regular users as detailed in Section~\ref{sec/dataset}. 

In Table~\ref{t:evasion_new}, results of $D_1'$ and $D_3'$ show that \model{}'s performance significantly improved after updating the classifier with only 3\% added attackers.
Notably, we observe a pronounced increase in recall, indicating that \model{} is capable of detecting new types of drainers with only limited examples of such attackers. 
\diff{
These results confirm that \model{} can effectively capture their hidden complex relationships within two graphs, which is difficult to change for attackers.
However, the precision for $D_3'$ remains suboptimal.
}
In the next section, we will discuss ways to make \model{} more robust against new types of drainers.

% To prevent evaluating on users seen in training, we train only on users that went inactive before a cutoff date.
% We re-trained the SVM for 5 different cutoff dates spaced apart by 2 days, and evaluated on the remaining data for each date.
% The number of active attackers at the end of the last day (August 10th) was 18(18.6\% of 97 drainers). 
% We trained and evaluated using a fixed ratio of drainers to regular users as previously described in Section~\ref{sec/dataset}. 
% Figure~\ref{fig:finetune} shows the evaluation results (F1-score) of experiments averaged over 5 runs.

% After just one update, with seven added attackers (7.2\%), \model{}'s performance significantly improved.
% As the model continues to update, its performance approaches the original level, and the performance difference between attack levels becomes small. 
% Attack 2 is noticeably more effective in evading the detector when more Ether is sent to victims.
% In the extreme scenario when 70\% of gifting is changed to buying ($L_4$) and 60\% of $Price_{avg}$ is sent, it becomes difficult to detect attackers even after updating.
% However, this is not likely to happen as it gives up a significant amount of profit from NFT draining. In the next section, we will discuss ways to make \model{} more robust against new types of drainers.

% \section{Case Study}
% \label{sec:casestudy}

% \subsection{False Positive}
% \diffnote{MR4}\diff{We found 422 false positives among 490,000 regular users.}

% \diffnote{MR5}\diff{\noindent\textbf{Detection of high-profile incidents:} sdf}

\section{Discussion \& Future Work}
\label{sec:discussion}

\diffnote{MR1}\diff{\subsection{Evasion attack}
\label{disc_evasion}
For evasion attacks against \model{}, the main limitation of our approach lies in the case where the attackers mask their liquidation process through multiple sales using multiple accounts.
Still, this type of attack yields a high trading volume of an NFT in a very short time and sometimes results in cycles in the NFT transaction graph.
Thus, this can be distinguished from regular trades.
\diffnote{R3C2} To detect this type of attacker, \model{} can integrate with other protection systems, such as a wash trading detection system. 
It can capture their irregular trades by 1) detecting a rapid trade sequence using transaction velocity~\cite{von2022nft} or 2) finding cyclic trading patterns between users~\cite{von2022nft, das2021understanding}.

Furthermore, we note that with a simple defense strategy, \model{} can detect evasive attackers that use a single account, but it still shows low precision.
This low precision comes from the fact that they modified trading patterns to mimic regular users, and we re-train the model using them.
To solve this problem, it is important to identify new characteristics of drainers that they cannot alter.
We manually analyzed the blockchain transaction history and found that drainers tend to receive (steal) multiple NFTs from each of their victims nearly simultaneously.
This phenomenon is attributable to the drainers' strategy; they distribute phishing websites, aiming for maximum reach and victim count, which leads to a synchronous influx of victims.
However, drainers cannot change this pattern since they lack control over the moment each victim is compromised.
Using this insight, we can consider integrating this characteristic into \model{}.
Specifically, in the User Graph, we can add \textit{interaction timespan} and \textit{the number of NFTs traded} between users to edge attributes.
In the NFT-User graph, we can add \textit{blockchain timestamp} when the transaction occurs to the NFT ownership edge attributes.  
This information can help \model{} better distinguish evasive attackers from regular users.

% The defense strategy can be improved by updating the transaction context extractor and social context extractor, not only the SVM.
% Furthermore, additional adversarial examples could be created to augment training data.
% In Section~\ref{sec:defense}, we observe that the performance increases significantly after the detector is updated with only a few adversarial attackers. 
% The detection performance will recover quickly by updating \model{} with examples similar to new drainers.

}

\begin{figure}
\diffnote{MR2}
\ifdiff
\begin{mdframed}[backgroundcolor=blue!10]
\fi
    \centering
    \includegraphics[width=0.9\columnwidth]{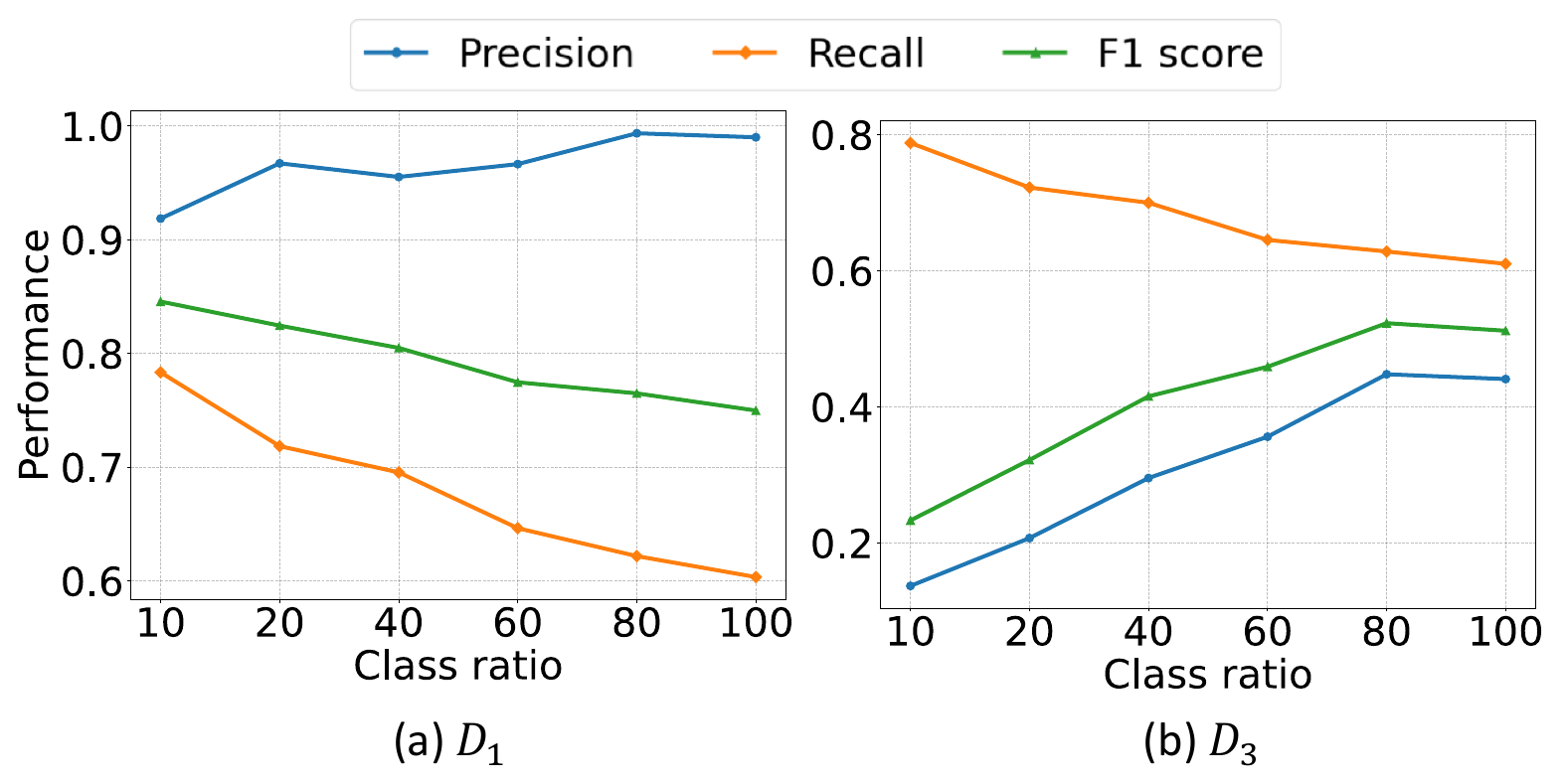}
      \vspace{-0.2cm}
    \caption{\diff{Detection results for $D_1$ and $D_3$ datasets, based on the class ratio in the training dataset. The X-axis indicates the relative number of regular users to drainers.}
}
    % The active attackers until each date are used for additional training, and the remains are used for evaluation. For each evaluation dataset, }
    \vspace{-0.2cm}
    \label{fig:size_eval}

\ifdiff
\end{mdframed}
\fi

\vspace{-0.3cm}
\end{figure}
\diffnote{MR2} \diff{\subsection{Influence of training class ratio on performance} 
\label{sec:disc_classratio}
We trained our model to prioritize F1-score, a suitable metric for imbalanced datasets.
However, the real-world significance of spotting unidentified drainers underscores the need for high recall. 
It should be noted that \model{} can be tailored for high recall by altering the ratio of drainers to regular users in the training dataset.

Figure~\ref{fig:size_eval} shows the influence of class ratio in the training dataset on overall performance. 
Given the limited reported drainer accounts available, we can increase the number of regular users. 
Within the regular user category in each dataset, we fixed the number of heavy regular users as ten times that of the drainers.
\model{} trained with fewer regular users yields higher recall, facilitating the detection of previously undetected drainers. 
Conversely, as we incorporate more regular users into training, \model{} learns from a more diverse set, increasing precision. 
However, using a large number of regular users leads to a high-class imbalance in training, causing the model to focus more on avoiding false positives than on detecting drainers, which decreases recall. 
To address this limitation, we can expand drainer samples used for training by employing synthetic minority over-sampling techniques.
This approach holds promise for enhancing both recall and precision.}

\diffnote{MR2}\diff{\subsection{Usage scenarios of \model{}} 
\label{sec:disc_scenario}
 \model{} operates on a robust foundation consisting of a database enriched with Ethereum transaction data and a curated list of identified drainer accounts. This database undergoes real-time updates, fetching the latest transaction data directly from the blockchain.
\diffnotenext{R3C2}
Also, it augments its drainer account list by integrating victim reports and phishing website detection systems. 
By analyzing data from phishing website detection systems, we can also collect on drainer accounts that have not yet started draining.
Each time a transaction occurs, \model{} updates the profiles of the involved users and corresponding NFTs. The system can scan multiple targets simultaneously, assigning risk scores to each user profile. These evaluations are essential for enhancing the security measures of software crypto wallets and NFT marketplaces.

Specifically, \model{} can integrate with software crypto wallets—tools that facilitate interactions with blockchains. 
The following scenario can stop the draining that caused by victims signing transactions with abused contracts.
When a user tries to sign a transaction, \model{} could be used to cross-reference the recipient’s account against its drainer list. 
If the recipient is on the drainer list, the transaction is promptly halted, and the user is warned of this information. 
Otherwise, the risk score attached to the account is checked. Accounts with risk scores surpassing administrator-set thresholds trigger warnings to the user, prompting them to proceed with or abandon the transaction. 
This real-time decision is instrumental in preventing potential victimization.

On the other hand, marketplace administrators are notified with real-time updates on the drainer list and user risk scores. They can employ an automated banning mechanism that relies on high risk score thresholds (which ensures high precision).
Alternatively, they can manually inspect accounts flagged by less stringent thresholds to increase recall. 
Once threats are confirmed, the marketplace can block the trading of stolen NFTs and notify affected users. The timely identification and banning of drainer accounts are important in undermining their revenue streams.

These administrators can subsequently incorporate newly identified drainer accounts into \model{}'s drainer list. If the count of such new additions surpasses a predetermined limit, \model{} can undergo a re-training phase using the expanded datasets to adapt and counteract the evolving strategies of drainers continually.

%market place에서는 
% In the current state of affairs, marketplaces can only take action on drainers through victim reports. Naturally, this causes a long delay between a draining event and moderators becoming aware, heightening the risk for additional victims.

  }

\diff{\subsection{Analysis of false positives \& negatives}
\label{sec:disc_fp}
% \diffnote{MR4} \model{} outperformed existing baselines in identifying drainer accounts. Nonetheless, it produced a number of false positives and false negatives. To enhance \model{}'s accuracy in the future, we delve into a case study of these inaccuracies.

\textbf{$\bullet$ False positive}: 
Of the 490,000 regular users evaluated (in $D_3$), 423 were flagged as drainers.
However, unreported drainers might be included in false positives.
To identify how many unreported drainers, \textit{potential drainers}, exist,
we manually verified each user using two indicators:
1) Connection with phishing attackers and 2) Possession of suspicious NFTs.
The detailed criteria refer to Appendix~\ref{apx:criteria}.

From our analysis, we identified 115 potential drainers.
Taking these into account, the adjusted performance metrics were 0.597 precision, 0.691 recall, and 0.641 F1 score within the D3 dataset.
We note that the actual number of unreported drainers could be higher.

One potential drainer sold 33 NFTs and gifted-out 19 NFTs from 79 gifted NFTs. 
Analyzing this user's Ethereum transaction history, we found records of Ether transfers with a reported phishing attacker that spanned 81 days.
Their prolonged interaction suggests a potential relationship, raising suspicions about the user's activities in the NFT space.

On the other hand, some regular users were misclassified as drainers because their legitimate trading behaviors resemble those of actual drainers.
We observed that the majority (88.8\%) of false positives acquired NFTs solely via gifting-in (instead of buying or minting) and quickly sold them within their short active timespan (See Appendix~\ref{apx:error_analysis}). For instance, one user sold 253 NFTs, from which 241 were recieved as gifts, in just five days. Intriguingly, the user received 233 NFTs from another account over two days and swiftly sold most of them. 

This behavior arises because sometimes individuals create several wallet accounts for better asset management to mitigate risks.
Such practices can mislead \model{} into incorrectly categorizing them as drainers, especially if they rapidly sell NFTs after receiving them. However, a distinguishable pattern exists: during a draining attack, all NFTs of a victim are instantly transferred to the drainer, whereas benign users transfer their NFTs across multiple hours or days. Our model, unfortunately, overlooked this temporal distinction in user interactions. It's crucial to factor in the duration over which interactions occur between users in future refinements.

\textbf{$\bullet$ False negative}: We analyze drainers misclassified as regular users. 
False negatives have a lower out-in ratio than other drainers; 46.5\% of them never sold an NFT, and 43.2\% of them never gifted out an NFT (Appendix~\ref{apx:error_analysis}).
A key characteristic of drainers comes from when they liquidate or transfer the stolen NFTs to affiliated users. It seems that \model{} failed to detect them due to the lack of such processes.

}

\begin{figure}
\diffnote{MR5}
\ifdiff
\begin{mdframed}[backgroundcolor=blue!10]
\fi
    \centering
    \includegraphics[width=0.8\columnwidth]{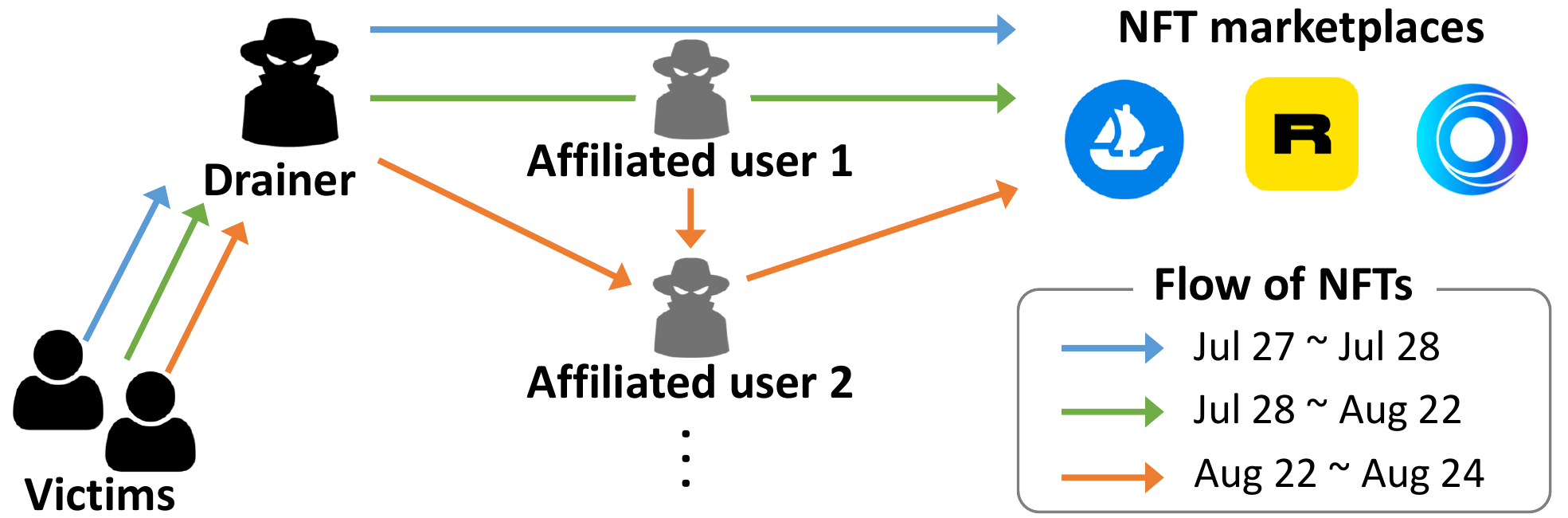}
      \vspace{-0.1cm}
    \caption{\diff{Trading pattern of a drainer related to North Korean threat actors. }
    \vspace{-0.3cm}
}
    \label{fig:casestudy}

\ifdiff
\end{mdframed}
\fi
\vspace{-0.3cm}
\end{figure}

\diffnote{MR5} \diff{\subsection{Analysis of high-profile incidents}
\model{} can detect drainers who conduct large-scale attacks.
We discuss high-profile incidents that made headlines in the media, all of which were detected by \model{}.

In December 2022, an incident attributed to North Korean state-sponsored threat actors notably garnered significant media attention~\cite{casestudy1_1,casestudy1_2,casestudy1_4}.
These attackers made off with digital assets worth thousands of dollars. 
The attackers set up nearly 500 decoy websites, including renowned NFT collection sites and marketplaces. 
One particular drainer stole 1,055 NFTs in total.
This actor exhibited a unique liquidation method, as illustrated in Figure~\ref{fig:casestudy}. 
On July 27, 2022, the NFTs stolen from victims began moving to this drainer's wallet. 
Over the next day, these stolen NFTs were swiftly offloaded on OpenSea. 
Starting July 28, newer stolen NFTs were transferred to an account, \textit{affiliated user 1}, which proceeded to sell them. 
By August 22, the drainer started directing newer stolen NFTs to a new account, \textit{affiliated user 2}, and any unsold NFTs of \textit{affiliated user 1} were also transferred to the latter. 
This shift suggests a strategic maneuver to employ a secondary affiliated account for continued sales, especially when the previous one neared detection. Cumulatively, they employed a network of 15 affiliated accounts to optimize their sales strategy.
Surprisingly, this account persisted in its draining activities and monetization until May 18, 2023. This persistence underscores the importance and timeliness of the real-time detection system.

Another noteworthy incident unfolded on October 21, 2022, which was started by the scammer known as \textit{Monkey Drainer}~\cite{casestudy2_1,casestudy2_2,casestudy2_3,casestudy2_4}.  
The drainer forged multiple accounts, mimicking influential Twitter accounts associated with the NFT community, such as those linked to the \textit{RTFKT} collection and \textit{Bored Ape Yacht Club} (BAYC) marketplace. 
They then disseminated posts that directed users to counterfeit NFT websites, baiting them with the promise of rewards or benefits. 
Over four days, assets amounting to roughly \$3.5 million, including 251 NFTs, were stolen. 
After draining, the stolen NFTs were quickly shifted to four associated users, who in turn sold them shortly after acquisition.

% After that, stolen NFTs were sent to affiliated users and sold.
% 처음에는 하나의 계정에게 보내고 그 계정이 판매.
% 어느 순간 새로운 계정을 생성한 뒤 
% 처음 하루동안은 훔친 NFT를 판매했고 
% 그 뒤에는 총 15개의 affiliated user에게 전달한 뒤 판판매했다.
% 1024 giftin
% 1017 giftout

% 7월 27일 처음 활동
% 5월 23년 마지막 활동

}

\section{Related Work}
% In this section, we briefly review previous work related to scams in the NFT ecosystem. Next, we discuss prior work on detecting Ethereum phishing accounts and briefly review graph neural networks, which is the key part of our detection system.

\noindent\textbf{Suspicious behaviors in NFT markets:} With the increasing popularity of NFTs, suspicious activities targeting NFTs are also rising.
A few studies exist for analyzing security issues in the NFT ecosystem, such as wash tradings and shill bidding. However, to the best of our knowledge, we are the first to perform an in-depth study of NFT drainers and propose an NFT drainer detection system. 

Das et al.~\cite{das2021understanding} conducted a comprehensive study of design weaknesses originating from the
NFT marketplaces and external entities. Also, they investigated various types of fraudulent user activities occurring in NFT marketplaces, such as counterfeit NFT creation, wash trading, and shill bidding.
Von et al.~\cite{von2022nft} quantified market abuse in the NFT ecosystem with their proposed NFT wash trading detection algorithm.
Roy et al.~\cite{roy2023demystifying} conducted a longitudinal analysis of Twitter accounts that consistently promote fraudulent NFT collections through giveaway competitions and NFT phishing attacks.

\noindent\textbf{Ethereum phishing scam detection:} Ethereum phishing scam detection can be categorized into two main types: feature-based and graph-based approaches. Chen et al.~\cite{chen2020phishing} extracted 119-dimensional statistical features to consider the 1-order neighbors of the node as well as the node itself. They used a LightGBM-based dual-sampling ensemble algorithm to classify phishing nodes.

Another line of research focuses on network representation.
Wu et al.~\cite{wu2020phishers} proposed Trans2Vec, which is a modified random walk-based network embedding method with biases of transaction amount and timestamp for neighbor sampling.
Chen et al.~\cite{EGCN} introduced E-GCN, the first Ethereum phishing scam detection method based on Graph Neural Networks (GNN). They extracted 8-dimensional statistical features, such as in/out-degree, number of neighbors, etc., and fed them into a GCN for embedding.
Li et al.~\cite{li2022ttagn} constructed edge representations from transaction records to capture the temporal relationship between users. The edge representations are aggregated into node representations and used to obtain structural features using GCN.
Unlike the above works that approached using node classification, Zhang et al.~\cite{zhang2021blockchain} regarded the problem as a graph classification. They used hierarchical graph pooling layers to extract node-level representations, which were then aggregated to form graph-level representations.

However, the works discussed above are difficult to apply to NFT drainers detection due to the characteristics of the NFT ecosystem, as mentioned in Section \ref{s:background}. 

% Furthermore, they construct homogeneous graph where nodes and edges are users and transaction between them,  
% Especially the process of liquidation of stolen NFTs is distinguishable from regular transactions of each NFT. 

% The study on detecting using GNN do not consider transaction type (\textit{sale, transfer}) between traders, which makes it hard to capture their relationships in the NFT trading. 
% Thus, discovering distinctive behavior traits of NFT phishing accounts and designing a detection system with them is in need.
%기존의 이더리움 피싱은 대부분 cryptocurrency를 타겟으로 하기 때문에 그에 맞춰져있다.
%NFT는 각각 Unique하기 때문에 각 NFT의 transaction 기록과 
%현금화의 방법이 다르기 때문에 다른 structure을 가질 것 
%NFT marking이 가능함에 의해 나타나는 behavior 특징 --> NFT 각자를 고려 
%NFT는 tx type이 여러개, 청산을 직접 해야함, 훔친 NFT를 tracking 할 수 있기 때문에 

\noindent\textbf{Graph Neural Network:} In recent years, deep learning methods have achieved remarkable performance in various fields. Deep neural networks have also been applied to graph data to leverage the structural properties of graphs.

Graph Convolution Networks (GCNs)~\cite{kipf2016semi} is one of the most prominent graph neural network models. GCNs perform convolution operations on graph data and learn embeddings of nodes by aggregating features from neighboring nodes.
Unlike GCNs, which uses information from adjacent nodes as is, Graph Attention Networks (GATs)~\cite{GAT} utilize information from neighbors by using node attention. 
Multi-head attention is used to learn a number of attentions, and the node features obtained from each attention are concatenated to form a single feature. 
GraphSAGE~\cite{graphsage} is an inductive GNN model, which generalizes the unobserved nodes. 
By incorporating node features into the learning algorithm and aggregator functions, it can learn the distribution of neighboring node features and the topological structure for the neighbors of each node.

\section{Conclusion}
NFT stealers are a significant threat to the NFT trading ecosystem.
Despite the increasing damages caused by NFT drainers, their behaviors are not well studied. 
To conduct an in-depth study on NFT drainers, we construct NFT phishing scam datasets.
We verify that they have different transaction contexts and social contexts compared to regular users.
Based on our measurement results, we propose a detection model, \model{}, tailored to detect drainers in the NFT environment. 
\model{} is able to generate a user representation that considers NFT transaction context and social context. 
Evaluated on real-world NFT transaction data, we verify our model's effectiveness and robustness.
We believe that our findings and detection method will contribute to the security NFT ecosystem.
% \balance

\clearpage
\bibliographystyle{IEEEtranS}
\bibliography{reference}
% \appendix

\clearpage
% \newpage
\begin{appendices}
% \diff{
% \section{Scenarios in which gifting happens}
% \label{apx:gift_ex}
% In the NFT realm, there are various scenarios in which gifting happens. For instance, a user might receive an overwhelming amount of unsolicited NFT gifts, commonly known as spam NFT gifting. 
% To combat this, the user might establish a new wallet and meticulously transfer only the desired NFTs from their old wallet.
% Furthermore, in blockchain-based games that incorporate NFTs, there exist specific smart contracts that facilitate the gifting of these NFTs to other players or accounts. 
% As another example, gifts can be used between users to avoid monitoring when manipulating markets by wash trading~\cite{das2021understanding}.
% This mechanism can occasionally be exploited by users to skirt around monitoring systems, especially when attempting market manipulation tactics like wash trading.}

\diff{
\section{Implementation Details}
\label{apx:implementation}
The embedding size of both \firstmodule{} (TCE) and \secondmodule{} is set to 64, and the learning rate is set to 6e-4 and 2e-3, respectively. 
In TCE, the number of attention heads is set to 8. 
The regularization parameter and gamma of SVM are set to 0.1 and 0.1, respectively.
As the classifier for each baseline, we choose the one with better performance between SVM and lightGBM~\cite{ke2017lightgbm}, another machine learning algorithm widely used for classification problems. 
Feature-based methods, E-GCN, and N-GCN use lightGBM, and the rest use SVM.
}

\diff{
\section{Criteria of potential drainer}
\label{apx:criteria}
% To identify how many \textit{potential drainers} exist, we manually verified each user based on the following criteria:

\noindent C1. Users possessing \textit{suspicious NFTs} that were banned from trading on OpenSea due to suspicious activities.

\noindent C2. Users who have consistently gifted NFTs to another account, where the receiving account holds \textit{suspicious NFTs}.

\noindent C3. Users who engaged in multiple Ether or NFT gifts over time with accounts labeled as phishing attackers on Etherscan.

\noindent C4. Users who are newly reported.

We note that the actual number of potential drainers could be higher, given that our assessment 1) did not account for NFTs that users no longer owned in C1 \& C2 and 2) only considered phishing attackers as tagged by Etherscan in C3.
}

\section{Distribution of transactions}
\label{apx:tx_dist}

\begin{figure}[ht]
    \centering
    \includegraphics[width=0.45\textwidth]{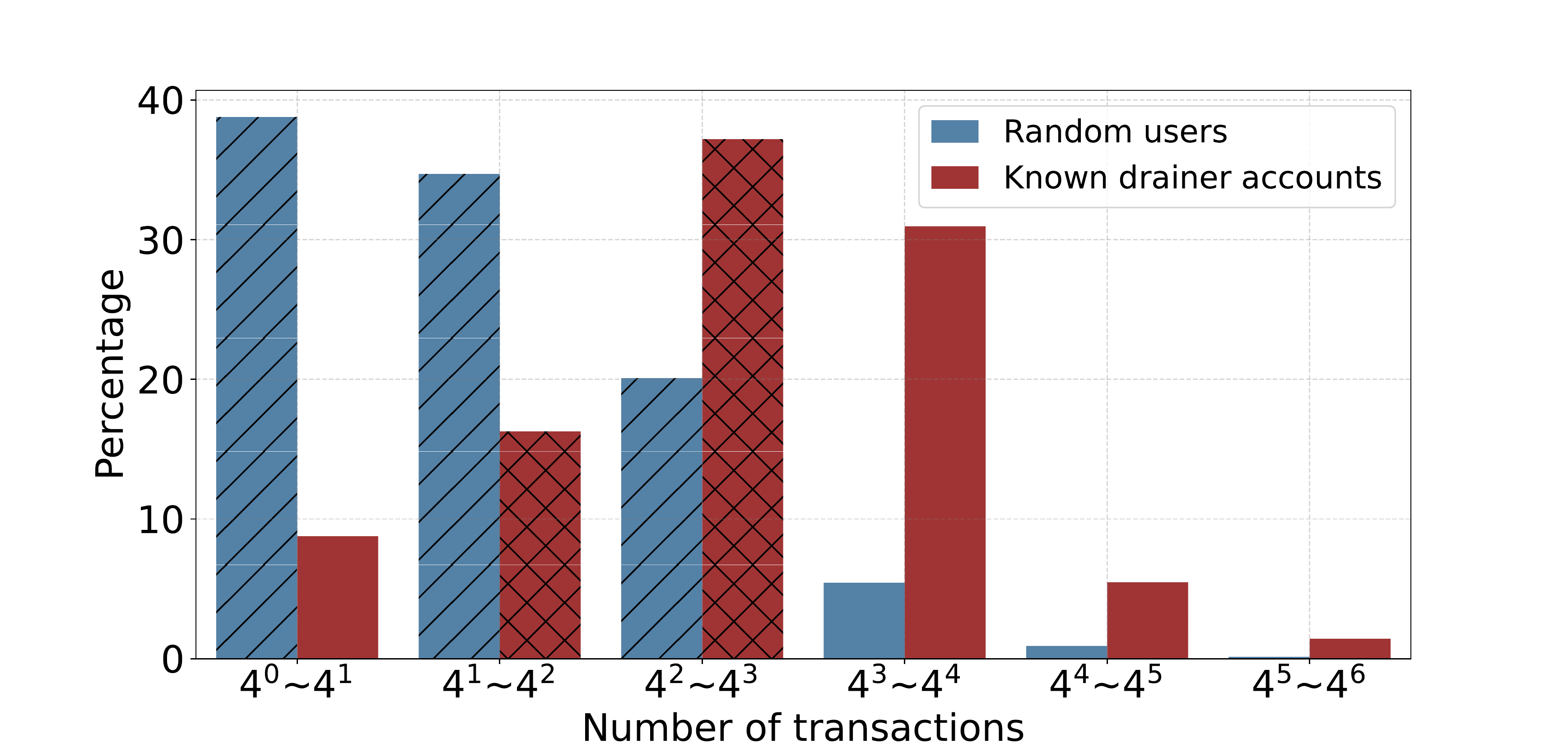}
    \caption{Distribution of random users' transactions and drainers' transactions. }
    \label{fig:tx_dist}
    
\end{figure}

Figure~\ref{fig:tx_dist} shows the distribution of random users' transactions and drainers' transactions.
We analyze drainers and random users who were active between January 1, 2022, and July 31, 2022.
Drainers and sampled random users show noticeably different distributions of transactions. 
In this paper, we define heavy users as accounts that are involved in transactions of more than 50.

\section{Feature analysis}
\label{apx:feat_analysis}
To better understand the trading behavior of different types of users, we analyze the activities of 645 drainers, 637 affiliated users, and a sample of 10,000 regular users who were active between January 1st and August 31st, 2022.
We compare these three groups by plotting the cumulative distribution functions (CDFs) for 19 different dimensions.
To make these visualizations easier to interpret, we limit the x-axis of each graph to 100.
We can observe that drainers exhibit distinct behavior patterns compared to regular users. Furthermore, we find that affiliated users also show different patterns of behavior, which sets them apart from both regular users and drainers.
  
\subsection{Number of transactions}
\label{apx:feat_analysis_1}
\begin{figure}[ht]
    \centering
    \includegraphics[width=0.45\textwidth]{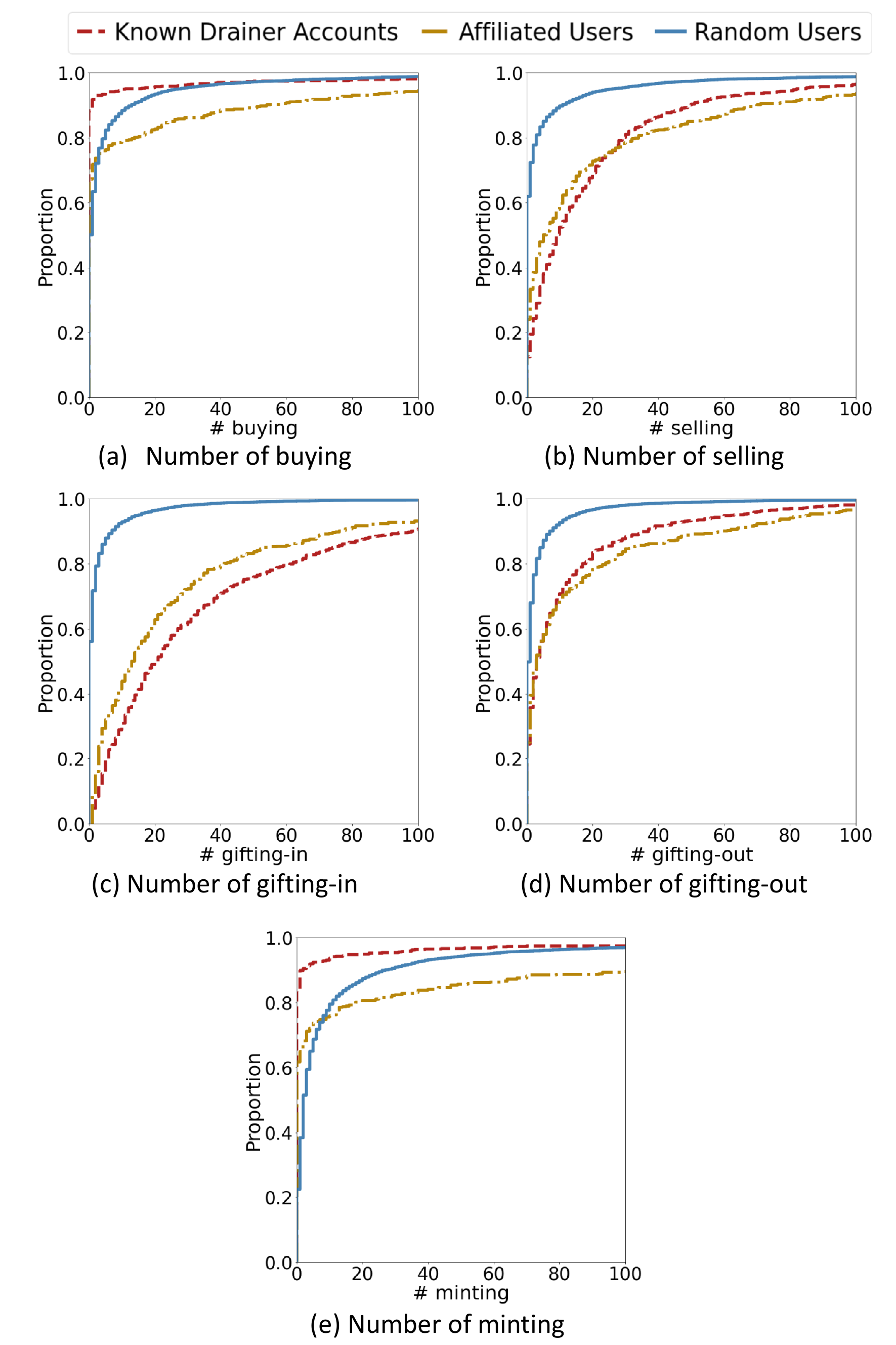}
    \caption{CDFs of the number of transactions according to the user types.}
    \label{f:feat1}
\end{figure}
In this section, we focus on the number of transactions users made for each transaction type. 
Figure~\ref{f:feat1} illustrates CDFs of the number of each transaction type: (a) buying, (b) selling, (c) gifting-in, (d) gifting-out, and (e) minting.

It is apparent that drainers are more active in selling, gifting-in, and gifting-out transactions. However, they rarely participate in buying and minting transactions when compared to regular users. These results suggest that the primary focus of drainers in the NFT ecosystem is on draining NFTs. 

Upon closer examination, trends between drainers and affiliated users show little difference in selling, gifting-in, and gifting-out transactions. Interestingly, affiliated users are more active in buying and minting NFTs, unlike drainers. This behavior sets them apart from both drainers and regular users.

\subsection{Number of collections}
\label{apx:feat_analysis_2}
\begin{figure}[ht]
    \centering
    \includegraphics[width=0.45\textwidth]{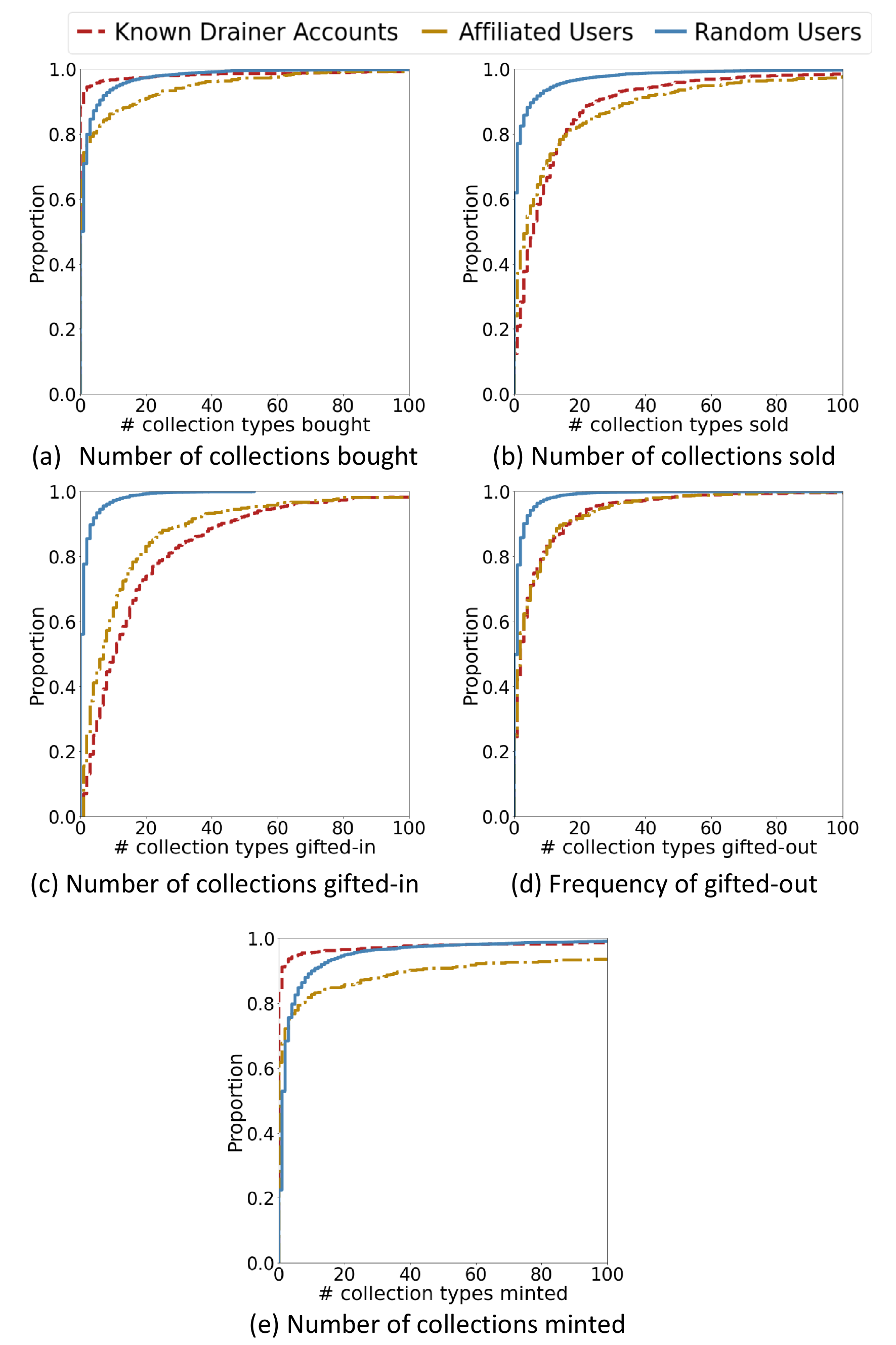}
    \caption{CDFs of the number of collections according to the user types.}
    \label{f:feat2}
\end{figure}

Figure~\ref{f:feat2} illustrates CDFs of the number of collections for each transaction type: (a) buying, (b) selling, (c) gifting-in, (d) gifting-out, and (e) minting.

We find that the number of collections for each transaction type is generally smaller than the number of transactions. 
It is well known that NFT users tend to form communities based on specific collections and trade within those communities~\cite{nadini2021mapping}. 
Although drainers have different intentions from regular users, they also trade a smaller number of collections than transactions.
This is because they steal NFTs that are collected by regular users.
However, due to their high levels of gifting-in, gifting-out, and selling transactions, drainers have a greater diversity of collections in those three types of transactions. 
As a result, they exhibit a significant difference from regular users in the number of NFT collections gifted-in, gifted-out, and sold.

In contrast, affiliated users are observed to actively participate in trading a wide range of collections across all types of transactions. This is because they are actively engaged in all types of transactions.

\subsection{Number of neighbors}
\label{apx:feat_analysis_3}
\begin{figure}[ht]
    \centering
\includegraphics[width=0.45\textwidth]{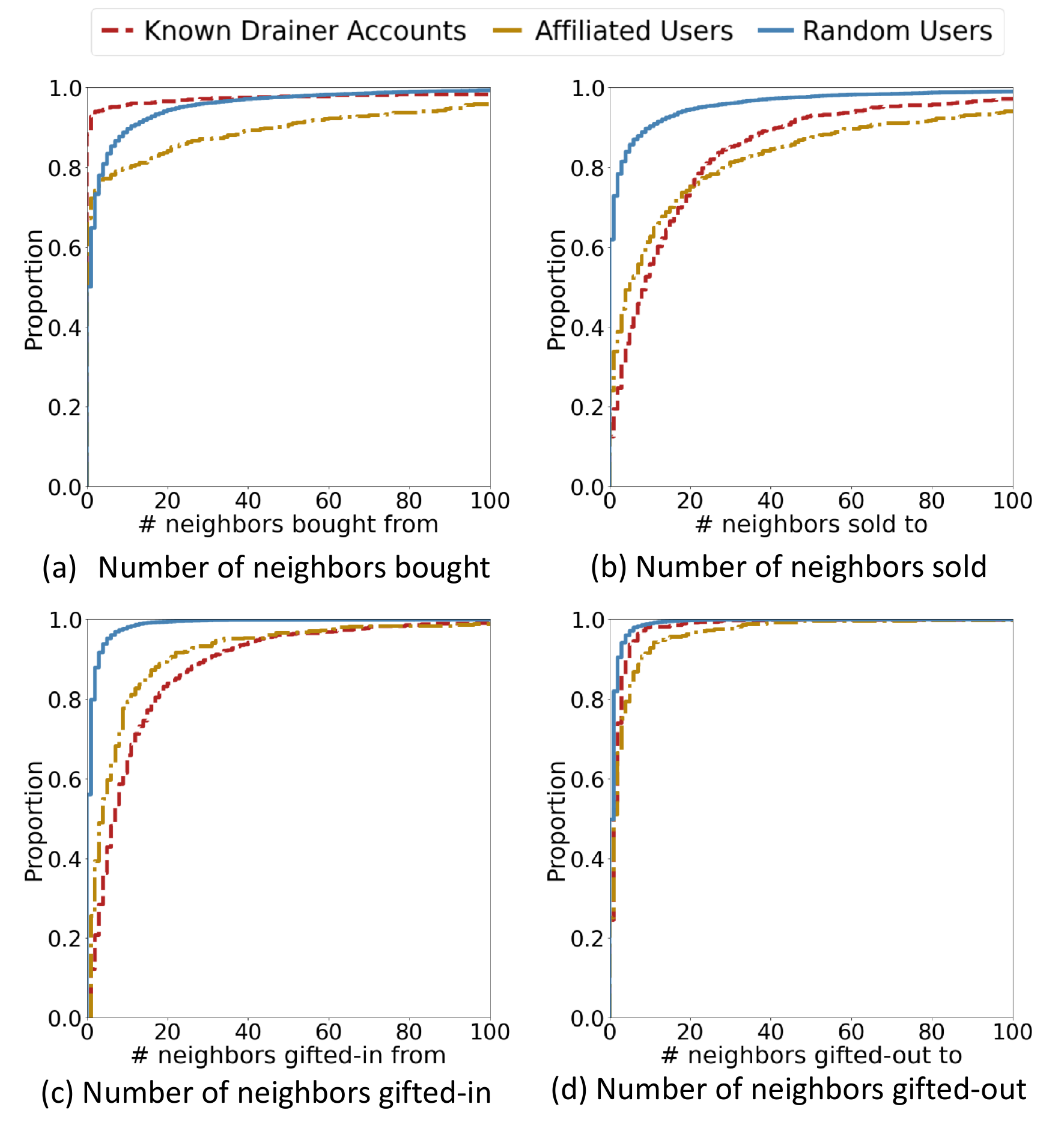}
    \caption{CDFs of the number of neighbors according to the user types.}
    \label{f:feat3}
\end{figure}

 Figure~\ref{f:feat3} illustrates CDFs of the number of neighbors for each transaction type: (a) buying, (b) selling, (c) gifting-in, and (d) gifting-out.
 
We analyze the number of neighbors a user has for each transaction type by considering the accounts with which a user has made a transaction as their neighbors. 
For buying and selling transactions, the distribution of neighbors is similar to that of the transactions themselves. 
However, for gifting-in and gifting-out transactions, the distribution of neighbors is significantly different. 
Most users gift NFTs to only a few neighbors, while they sell or buy NFTs with many neighbors. 
This suggests that NFT users have specific relationships through gifting NFTs, given that gifting is a process of transferring ownership without any payment.

This phenomenon is more pronounced in drainers than in regular users. When drainers steal NFTs, they may acquire all the tokens from each victim, resulting in a smaller number of gifting-in neighbors than the number of NFTs gifted-in. 
Additionally, drainers only gift NFTs to a few affiliated users, resulting in a much smaller number of gifting-out neighbors than the number of NFTs gifted-out.

In the case of affiliated users, their distributions of gifting-in and gifting-out are as expected, similar to those of drainers.

\subsection{Ratio \& Frequency \& Active timespan}
\label{apx:feat_analysis_4}
\begin{figure}[ht]
    \centering
\includegraphics[width=0.45\textwidth]{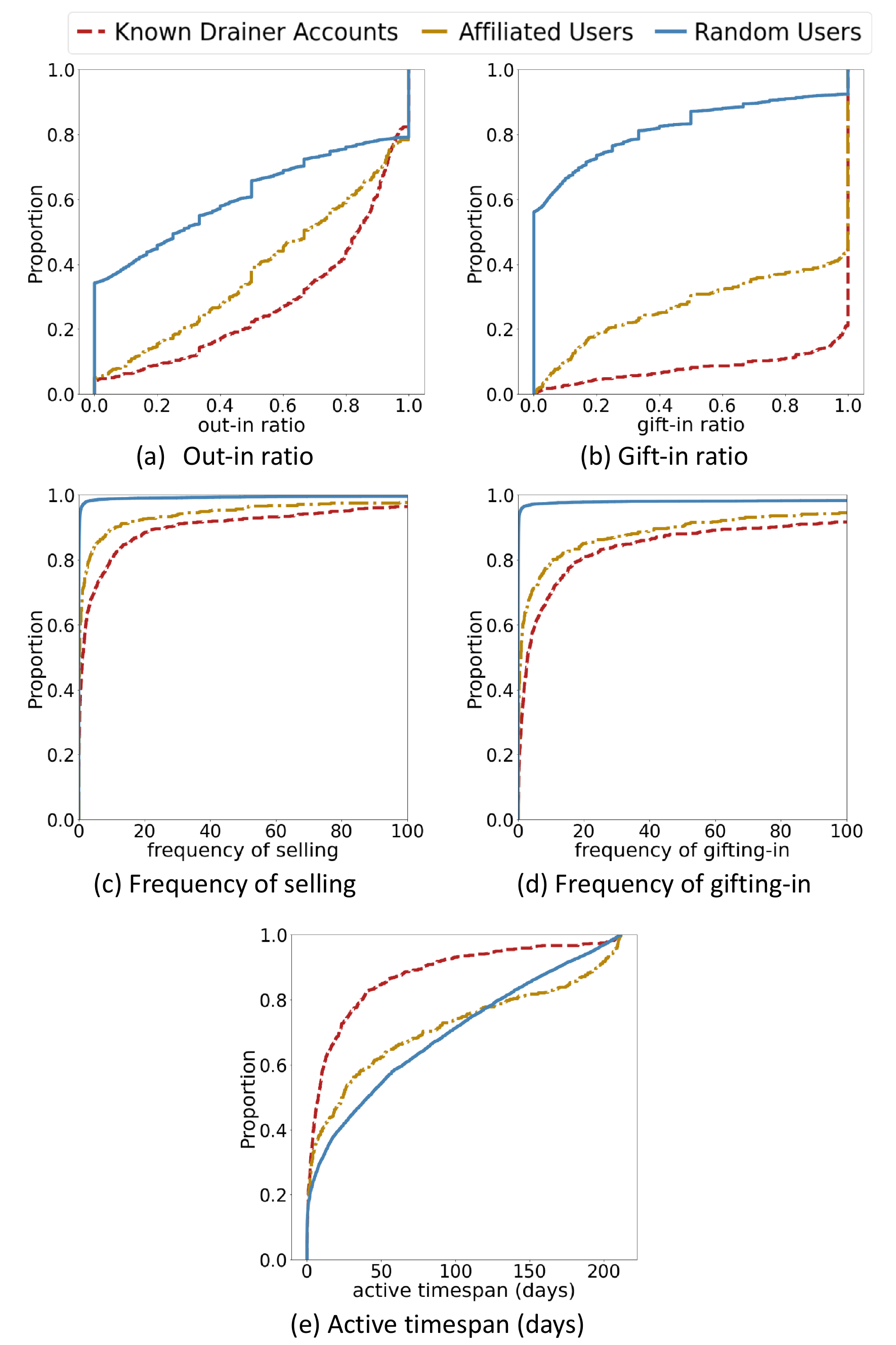}
    \caption{CDFs of gift-in ratio, out-in ratio, frequency of sale, frequency of gift-in, and active timespan according to the user types.}
    \label{f:feat4}
\end{figure}

Figure~\ref{f:feat4} illustrates CDFs of (a) out-in ratio, (b) gift-in ratio, (c) frequency of selling, (d) frequency of gifting-in, and (e) active timespan.
 
Drainers have a higher gift-in ratio than regular users and affiliated users. This is because they do not engage in buying and minting NFTs, but rather steal a large number of NFTs in a short period of time. This results in a high frequency of gifting-in transactions.
Also, drainers are more likely to transfer out their NFTs than regular users. They also sell their NFT much more frequently in a short active timespan. 

The behavior of affiliated users falls between that of drainers and regular users. They are active in all types of transactions, particularly gifting-in, which results in a high gift-in ratio similar to drainers. 
However, their active period is not as short as drainers.

% \onecolumn
\section{Ablation study results on $D_2$}
\label{apx:ab_study}

\begin{table}[ht]
\caption{
The results of ablation experiments averaged over 5 runs. Removing \textit{relation in SCE} refers to the modified version of \model{} where the representation does not consider transaction types between users in the user graph.}
\centering
\ra{1.1}
\footnotesize
% \resizebox{\columnwidth}{!}{%
\begin{tabular}{l ccc}
    \toprule 
    Dataset (ratio)  
    & \multicolumn{3}{c}{$D_2$ (1:100)}  \\
    \midrule
    Removed & Pre. & Rec. & F1  \\
    \midrule 
    User features &
    0.864 & 0.599 & 0.707  \\
    NFT transaction context &
    0.878 & 0.585 & 0.702 \\
    Social context&
    0.853 & 0.591 & 0.698 \\
    Relation in SCE &
    0.817 & 0.589 & 0.684  \\
    \midrule
    \model{} &
    0.878 & 0.621 & 0.727 \\
    \bottomrule
\end{tabular}
% }
\end{table}

\section{Evasion attack results on $D_2$}
\label{apx:evasion_attack}
\begin{table}[ht]
\diffnote{MR1}
\ifdiff
\begin{mdframed}[backgroundcolor=blue!10]
\fi
\caption{The results of evasion attacks on dataset $D_2$ averaged over 5 runs.}
\centering
\ra{1}
\footnotesize
% \begin{tabular}{ccr c@{\hspace{5pt}}c@{\hspace{5pt}}c| c@{\hspace{5pt}}c@{\hspace{5pt}}c }
\begin{tabular}{ccr ccc| ccc }
    \toprule
    & \multicolumn{2}{c}{\makecell[c]{Dataset\\(ratio)}}
    & \multicolumn{3}{c}{$D_2$ (1:100)} 
    & \multicolumn{3}{c}{$D_2'$ (1:100)}\\
    \cmidrule{2-9}
    & L & X
    & Pre. & Rec. & F1 
    & Pre. & Rec. & F1 \\
    \midrule
    \multirow{3}{*}{Attack1}
    & 10 &&
    0.771 & 0.481 & 0.593 &
    0.767 & 0.541 & 0.635 \\
    & 30 & N/A&
    0.730 & 0.386 & 0.505 &
    0.758 & 0.509 & 0.609 \\
    & 50 && 
    0.712 & 0.354 & 0.472 &
    0.754 & 0.499 & 0.600 \\
    \midrule
    \multirow{3}{*}{Attack2}
    & 10 &&
    0.685 & 0.347 & 0.461 &
    0.774 & 0.603 & 0.677 \\
    & 30 & N/A&
    0.546 & 0.192 & 0.284 &
    0.779 & 0.640 & 0.703 \\
    & 50 && 
    0.466 & 0.139 & 0.214 &
    0.788 & 0.666 & 0.722 \\
    \midrule
    \multirow{9}{*}{Attack3}
    & \multirow{3}{*}{10} 
    & 1 &
    0.427 & 0.117 & 0.183 &
    0.732 & 0.600 & 0.659 \\
    && 10&
    0.416 & 0.113 & 0.178 &
    0.728 & 0.583 & 0.647 \\
    && 60 &
    0.414 & 0.111 & 0.175 & 
    0.722 & 0.573 & 0.638 \\
    \cmidrule{2-9}
    & \multirow{3}{*}{30} 
    & 1 &
    0.446 & 0.127 & 0.197 & 
    0.752 & 0.663 & 0.704 \\
    && 10 &
    0.416 & 0.113 & 0.177 &
    0.741 & 0.638 & 0.684 \\
    && 60 &
    0.386 & 0.099 & 0.157 &
    0.721 & 0.625 & 0.669 \\
    \cmidrule{2-9}
    & \multirow{3}{*}{50} 
    & 1 &
    0.499 & 0.156 & 0.237 &
    0.775 & 0.686 & 0.727 \\
    && 10&
    0.456 & 0.133 & 0.205 &
    0.776 & 0.673 & 0.721 \\
    && 60 &
    0.420 & 0.114 & 0.180 &
    0.769 & 0.645 & 0.701 \\
    \midrule
    \multirow{9}{*}{Attack4}
    & \multirow{3}{*}{10} 
    & 1 &
    0.107 & 0.019 & 0.032 &
    0.650 & 0.425 & 0.513 \\
    && 10&
    0.107 & 0.019 & 0.032 &
    0.649 & 0.442 & 0.525 \\
    && 60 &
    0.117 & 0.021 & 0.036 &
    0.649 & 0.431 & 0.517 \\
    \cmidrule{2-9}
    & \multirow{3}{*}{30} 
    & 1 &
    0.108 & 0.019 & 0.032 &
    0.707 & 0.607 & 0.653 \\
    && 10 &
    0.107 & 0.019 & 0.032 &
    0.694 & 0.583 & 0.633 \\
    && 60 &
    0.074 & 0.012 & 0.021 &
    0.681 & 0.572 & 0.621 \\
    \cmidrule{2-9}
    & \multirow{3}{*}{50} 
    & 1 &
    0.074 & 0.012 & 0.021 &
    0.729 & 0.665 & 0.695 \\
    && 10&
    0.084 & 0.015 & 0.025 &
    0.715 & 0.650 & 0.681 \\
    && 60 &
    0.073 & 0.012 & 0.021 &
    0.712 & 0.638 & 0.673 \\
    \midrule
    \multicolumn{3}{c}{\model{}} & 
    \textbf{0.878} & \textbf{0.621} & \textbf{0.727} &
    \textbf{0.878} & \textbf{0.621} & \textbf{0.727} \\

    \bottomrule
\end{tabular}

\ifdiff
\end{mdframed}
\fi
\end{table}

\diff{
To evaluate \model{}'s detection capabilities against evasion tactics, we adjusted previous evaluation datasets.
These modifications were guided by the specific attack strategy and its attack level ($L$), where $L \in \{10, 30, 50\}$.
For Attack 1, we increased the number of minted NFTs by $L$\% of gifted-in NFTs. 
For Attack 2, the active timespan was extended by $L$\%.  
For Attack 3, we changed $L$\% of gifting-in transactions to buying transactions by sending X\% of the average sale price of each NFT to those victims, where $X \in \{1, 10, 60\}$. 
For Attack 4, we integrated the tactics of Attack 1 and Attack 2, both at level 50, with the methods of Attack 3.

We re-trained the classifier layer of \model{} (SVM) with 3\% of evasion attackers and evaluated the remaining evasion attackers on datasets $D_2'$.
}

\newpage
\onecolumn
\section{Evasion attack results with Varied Parameter $X$ on $D_1$ \& $D_3$}
\label{apx:evasion_attack_X}
\begin{table*}[h]
\diffnote{MR1}
\ifdiff
\begin{mdframed}[backgroundcolor=blue!10]
\fi
\caption{The results of evasion attacks on $D_1$ and $D_3$ datasets averaged over 5 runs. }
\label{t:evasion_X}
\centering
\ra{1}
\footnotesize
\begin{tabular}{ccr ccc| ccc | ccc | ccc}
    \toprule
    & Dataset (ratio) &
    & \multicolumn{3}{c}{$D_1$ (1:10)} 
    & \multicolumn{3}{c}{$D_1'$ (1:10)}
    & \multicolumn{3}{c}{$D_3$ (1:1000)}
    & \multicolumn{3}{c}{$D_3'$ (1:1000)}\\
    \cmidrule{2-15}
    & L & X
    & Pre. & Rec. & F1 
    & Pre. & Rec. & F1 
    & Pre. & Rec. & F1 
    & Pre. & Rec. & F1 \\
    \midrule
    \multirow{9}{*}{Attack3}
    & \multirow{3}{*}{10} 
    & 1 &
    0.872 & 0.116 & 0.205 &
    0.969 & 0.597 & 0.738 &
    0.084 & 0.120 & 0.099 &
    0.226 & 0.610 & 0.329 \\
    && 10&
    0.868 & 0.112 & 0.199 &
    0.966 & 0.582 & 0.725 &
    0.082 & 0.117 & 0.096 &
    0.222 & 0.595 & 0.323 \\
    && 60 &
    0.866 & 0.110 & 0.195 &
    0.966 & 0.574 & 0.719 &
    0.081 & 0.115 & 0.095 &
    0.220 & 0.591 & 0.320  \\
    \cmidrule{2-15}
    & \multirow{3}{*}{30} 
    & 1 &
    0.881 & 0.126 & 0.221 &
    0.970 & 0.659 & 0.784 &
    0.088 & 0.127 & 0.104 &
    0.236 & 0.681 & 0.350 \\
    && 10 &
    0.868 & 0.112 & 0.199 &
    0.968 & 0.634 & 0.766 &
    0.081 & 0.116 & 0.096 &
    0.231 & 0.653 & 0.341 \\
    && 60 &
    0.852 & 0.098 & 0.176 &
    0.965 & 0.625 & 0.758 &
    0.074 & 0.104 & 0.086 &
    0.222 & 0.635 & 0.328 \\
    \cmidrule{2-15}
    & \multirow{3}{*}{50} 
    & 1 &
    0.903 & 0.155 & 0.265 &
    0.972 & 0.684 & 0.802 &
    0.107 & 0.157 & 0.127 &
    0.265 & 0.704 & 0.384 \\
    && 10&
    0.889 & 0.133 & 0.231 &
    0.973 & 0.669 & 0.793 &
    0.093 & 0.133 & 0.110 &
    0.263 & 0.677 & 0.378 \\
    && 60 &
    0.873 & 0.114 & 0.202 &
    0.970 & 0.644 & 0.774 &
    0.082 & 0.118 & 0.097 &
    0.264 & 0.648 & 0.374 \\
    \midrule
    
    \multirow{9}{*}{Attack4}
    & \multirow{3}{*}{10} 
    & 1 &
    0.525 & 0.018 & 0.035 &
    0.953 & 0.421 & 0.583 &
    0.015 & 0.020 & 0.017 &
    0.170 & 0.477 & 0.251 \\
    && 10&
    0.525 & 0.018 & 0.035 &
    0.950 & 0.437 & 0.598 &
    0.016 & 0.021 & 0.018 &
    0.172 & 0.482 & 0.254 \\
    && 60 &
    0.551 & 0.020 & 0.039 &
    0.952 & 0.425 & 0.587 &
    0.017 & 0.023 & 0.019 &
    0.171 & 0.474 & 0.251 \\
    \cmidrule{2-15}
    & \multirow{3}{*}{30} 
    & 1 &
    0.525 & 0.018 & 0.035 &
    0.958 & 0.601 & 0.738 &
    0.016 & 0.021 & 0.018 &
    0.199 & 0.630 & 0.302\\
    && 10 &
    0.525 & 0.018 & 0.035 &
    0.955 & 0.576 & 0.718 &
    0.016 & 0.021 & 0.018 &
    0.192 & 0.610 & 0.292 \\
    && 60 &
    0.426 & 0.012 & 0.024 &
    0.956 & 0.563 & 0.709 &
    0.010 & 0.013 & 0.012 &
    0.183 & 0.587 & 0.278\\
    \cmidrule{2-15}
    & \multirow{3}{*}{50} 
    & 1 &
    0.430 & 0.012 & 0.024 &
    0.964 & 0.663 & 0.786 &
    0.011 & 0.014 & 0.012 &
    0.222 & 0.691 & 0.335 \\
    && 10&
    0.468 & 0.014 & 0.028 &
    0.961 & 0.642 & 0.770 &
    0.012 & 0.016 & 0.014 &
    0.211 & 0.658 & 0.319 \\
    && 60 &
    0.430 & 0.012 & 0.024 &
    0.961 & 0.634 & 0.764 &
    0.011 & 0.014 & 0.012 &
    0.207 & 0.651 & 0.314 \\
    \midrule
    \multicolumn{3}{c}{\model{}} & 
    \textbf{0.989} & \textbf{0.622} & \textbf{0.763} & 
    \textbf{0.989} & \textbf{0.622} & \textbf{0.763} & 
    \textbf{0.448} & \textbf{0.631} & \textbf{0.523} &
    \textbf{0.448} & \textbf{0.631} & \textbf{0.523} \\
    \bottomrule
\end{tabular}

\ifdiff
\end{mdframed}
\fi
\end{table*}

\diff{To evaluate \model{}'s detection capabilities against evasion tactics, we adjusted previous evaluation datasets.
These modifications were guided by the specific attack strategy and its attack level ($L$), where $L \in \{10, 30, 50\}$.
For Attack 3, we changed $L$\% of gifting-in transactions to buying transactions by sending X\% of the average sale price of each NFT to those victims, where $X \in \{1, 10, 60\}$. 
For Attack 4, we integrated the tactics of Attack 1 and Attack 2, both at level 50, with the methods of Attack 3. 

We re-trained the classifier layer of \model{} (SVM) with 3\% of evasion attackers and evaluated the remaining evasion attackers on datasets $D_1'$ and $D_3'$.}

\newpage
\section{Error Analysis}
\label{apx:error_analysis}
\begin{figure}[ht]
\diffnote{MR4}
\ifdiff
\begin{mdframed}[backgroundcolor=blue!10]
\fi
    \centering
    \includegraphics[width=0.6\columnwidth]{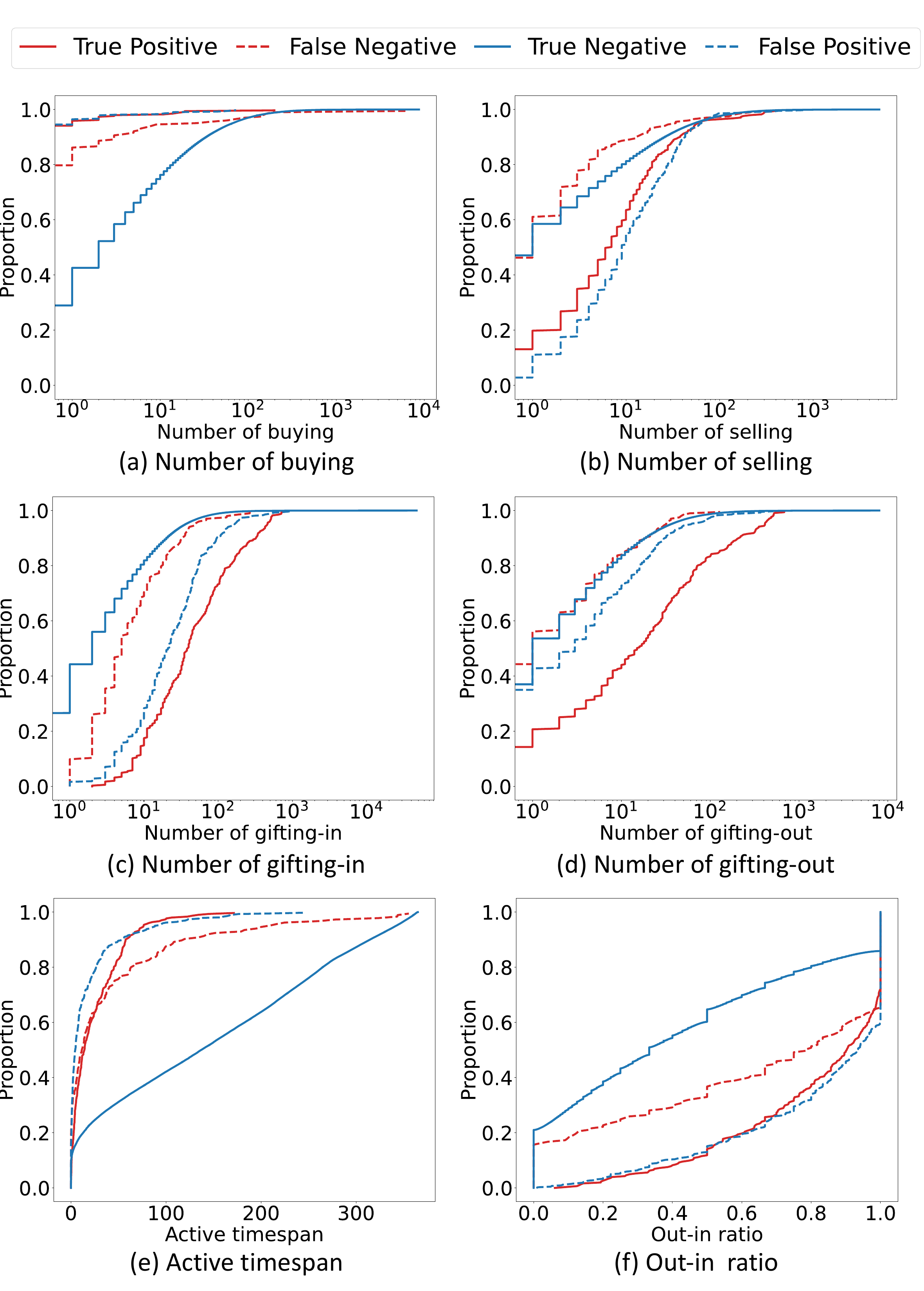}
      \vspace{-0.2cm}
    \caption{\diff{CDFs of gift-in ratio, out-in ratio, frequency of sale, frequency of gift-in, and active timespan according to the user
types.
}
}

\ifdiff
\end{mdframed}
\fi
\end{figure}

\end{appendices}

\end{document}